\documentclass[12pt,preprint]{aastex}
\usepackage{emulateapj5}
\begin{document}

\slugcomment{To appear in The Publications of the Astronomical Society
of the Pacific}

\title{Ultraviolet and optical properties of Narrow-Line
Seyfert 1 galaxies\altaffilmark{1}}

\altaffiltext{1}{Based on observations made with the NASA/ESA {\it
Hubble Space Telescope}, obtained from the data archive at the Space
Telescope Science Institute. STScI is operated by the Association of
Universities for Research in Astronomy, Inc., under NASA contract NAS
5-26555.}

\author{Anca Constantin and Joseph C. Shields} \affil{Department of
Physics and Astronomy, Ohio University, Athens, OH 45701;
constant@helios.phy.ohiou.edu}

\begin{abstract}

Narrow Line Seyfert 1 (NLS1) galaxies are remarkable for their extreme
continuum and emission line properties which are not well understood.
New results bearing on the spectroscopic characteristics of these
objects are presented here, with the aim of establishing their typical
ultraviolet (UV) and optical spectral behavior.  We employ {\it Hubble
Space Telescope} ({\it HST}) observations of 22 NLS1s, which represent
a substantial improvement over previous work in terms of data quality
and sample size.  High signal-to-noise (S/N) NLS1 composite spectra are
constructed, allowing accurate measurements of the continuum shape and
the strengths, ratios, and widths for lines, including weak features which
are barely identifiable in other Active Galactic Nuclei (AGN)
composites.  We find that the NLS1 sources have redder UV-blue
continua than those typically measured in other quasars and Seyferts.
Objects with UV line absorption show redder spectra, suggesting that
dust is important in modifying the continuum shapes.  The data also
permit a detailed investigation of the previously proposed link
between NLS1s and $z\ ^{>}_{\sim}\ 4$ quasars.  Direct comparison of
their composite spectra, as well as a Principal Component Analysis,
suggest that high-$z$ QSOs do not show a strong preference toward NLS1
behavior.

\end{abstract}


\section{Introduction}

Narrow-Line Seyfert 1 galaxies \citep{ost85, goo89} are a subclass of
AGNs that manifest a distinctive ensemble of properties.  They are
rather rare objects that exhibit relatively narrow broad lines, strong
\ion{Fe}{2} emission, and weak emission from the narrow line region;
they are more variable in X-rays than the Broad-Line Seyfert 1 objects
(``normal'' Seyfert 1s, hereafter Sy1), and exhibit pronounced soft
X-ray excesses.  They seem to cluster at one extreme end of the
\citet{bor92} ``Eigenvector 1'' (EV1) relation, as a result of their
tendency toward weak [\ion{O}{3}] $\lambda\lambda$4959, 5007 emission,
and narrow and blue-asymmetric H$\beta$ profiles.  Understanding this
EV1 is important for NLS1s in particular and for AGN in general since
it may be closely linked and possibly driven by the central engine
parameters, in particular $L/L_{Edd}$, the Eddington ratio
\citep{bor92, bor02}.  However, whether this is the main and the only
physical parameter that controls the NLS1 classification, with its
distinctive features, is still a matter of debate.

To date, NLS1 studies are built on either individual objects or on
samples for which the spectra span only narrow wavelength ranges;
also, with the exception of the X-ray observations, these samples are
small in general.  In particular, detailed investigations of the NLS1
blue/UV emission properties have been limited \citep{rod97, kur00}.
Studies of their spectral energy distributions (SEDs) that cover wide
bandpasses at a single epoch are almost completely missing.  This kind
of data is particularly useful for testing and constraining models
proposed for these sources.  Further examination of larger samples of
NLS1 emission spectra is clearly desirable.

Understanding the nature of NLS1s requires a detailed description of
their average behavior.  Therefore, a first goal of the present study
is to obtain a comprehensive spectral characterization of the typical
NLS1 galaxy.  The mounting number of high quality {\it HST} spectra of
NLS1 sources allows for a better definition of their spectral
properties in general, and their UV emission in particular.  In this
study, we make use of {\it HST} archival observations of 22 NLS1s, a
sample which is nearly twice as large as any of those used in previous
studies of the NLS1 ultraviolet line and continuum emission.  This
database includes several objects whose observations cover a wide
wavelength range (Ly$\alpha$ to H$\alpha$ region) permitting thus, for
the first time, a simultaneous survey of the NLS1 UV and optical
spectral features.  We construct average and median NLS1 composite
spectra and provide measurements of the resulting underlying
continuum, and the line strengths, ratios, and widths.

The second important question we attempt to address here is the degree
to which NLS1s and high redshift ($z \ga 4$) quasars share common
properties.  The possible connection between these two classes of
objects has been suggested by \citet{mat00}, who proposed a scenario
in which both NLS1s and high $z$ QSOs are in an early evolutionary
phase, such that accretion proceeds at or near the Eddington limit.
This analogy is mostly based on the indications for high metallicities
in both of these categories of sources, and on the presence of an
enhanced low-velocity component in the UV spectra of high redshift
quasars.  However, no direct comparison of their emission properties
has been attempted yet.  This is primarily due to their wide
separation in redshift, which makes it difficult to observe them in
the same spectral ranges.  The current availability of high-quality UV
spectra for the low redshift objects allows us to test the validity of
this picture.  We examine and compare the emission properties of the
NLS1s and $z \ga 4$ quasars, in order to determine the extent of the
similarities exhibited by these two classes of objects.  This comparison
provides additional insights into AGN behavior as a function of
redshift, luminosity, metallicity, and other physical parameters.

\section{The NLS1 sample \& data processing} \label{data}

Non-proprietary, archival {\it HST} Faint Object Spectrograph (FOS),
Goddard High Resolution Spectrograph (GHRS), and Space Telescope
Imaging Spectrograph (STIS) spectra of all objects known to us that
were previously identified in the literature as NLS1 galaxies (i.e.,
FWHM(H$\beta$) $\la$ 2000 km s$^{-1}$, $[$\ion{O}{3}]/H$\beta$ $<$ 3),
were retrieved from the Space Telescope Science Institute in the form
of calibrated data.  A number of these objects are borderline sources
in terms of this specific classification; however, they resemble the
NLS1 characteristics by every other definition, and they have been
extensively used in other NLS1 studies.  Table ~\ref{tbl-1} summarizes
the instrumental setup and the resulting total wavelength coverage
corresponding to each observation.  References relevant to the
classification of these objects are also listed.  Each line in the
table refers to observations obtained under a single observing
program; in all but one case (I Zw 1), these observation sets were
taken at a single epoch.  The data obtained under separate observing
programs are listed as different lines.

In general, observations of each object were acquired with multiple
gratings.  The full wavelength coverage was obtained by co-adding the
individual spectra, after resampling to a common dispersion (the
lowest number of \AA/pixel, to avoid loss of information), and
flux-averaging in the overlapping regions.  As indicated in Table
~\ref{tbl-1}, there are several galaxies for which multiple
observations were obtained at different epochs and in some cases with
different instruments.  Because these sources are characterized by
significant variability, the continuum level often needed minor
rescalings ($\la 10\%$) before averaging.  The use of multiple
instrumental configurations translated also into different spectral
resolutions $R$.  To obtain the final observed spectrum, it was
necessary to convert the individual observations first to a common
$R$, by gaussian smoothing of the spectra with high resolution, and
second, to a common dispersion.  For all objects, the averaging in the
common wavelength ranges was performed by weighting the spectra by the
reciprocal of their noise variance.

Many of these spectra display significant resonance line absorption.
The main focus of this analysis is on the emission lines, and
therefore, accurate measurements of these features required correction
for the intervening absorption.  When this appears as narrow lines
superimposed on either the emission lines or the continuum, we removed
the absorption feature through a simple interpolation.  When the
absorption was severe, especially near or within some of the strong
emission features (Ly$\alpha$, \ion{N}{5}, and \ion{C}{4}), a
conservative reconstruction of the line was attempted by low-order
polynomial fitting.  For these particular cases, as indicated in Table
~\ref{tbl-2}, the line flux and equivalent width (EW) measurements are
considered to be only lower limits.

The objects in this sample are subject to small amounts of Galactic
foreground extinction, spanning $A_V = 0.03 - 0.31$, with a typical
value of $\sim 0.05$ \citep{sch98}.  We corrected the spectra for the
resulting reddening using the empirical selective extinction function
of \citet{car89}.  No correction for intrinsic reddening was
attempted.

In order to combine the individual source observations into a single
composite spectrum, a proper alignment in wavelength of the emission
lines is necessary.  This requires, in turn, consistent measurements
of the redshift values $z$ used in the Doppler-correction process.
Therefore, we remeasured the redshifts using the \ion{C}{4}
$\lambda$1549 emission-line, which is measured in 21 out of 22 objects
(the \ion{C}{4} feature is a doublet $\lambda \lambda$1548.2, 1550.7,
with the simple average of $\lambda$1549.5, corresponding to the
optically thick case, consistent with photoionization models).  The
observed wavelength for each case was established by fitting a
Gaussian to the top 20\%\ of the profile.  For situations where the
measured line was significantly affected by absorption or low S/N, we
verified the result with fits to the top 50\%\ of the profile.  The
resulting $z$ determinations are in good agreement with previously
published values, with an estimated uncertainty of $\leq 0.001$, and
lead to a good superposition of the principal emission lines in the
rest-frame.  The redshift measurements obtained for each object from
\ion{C}{4} line fitting are listed in Table~\ref{tbl-2}.

Measurements of the emission features in the individual spectra were
carried out with line profile fitting. This was in general
unsuccessful when single Gaussians were used, as they tend to lose
flux from the wings if they are prominent, and from the peak if the
cores dominate.  The overall shapes of the lines measured in these
NLS1 spectra were generally well represented by single Lorentzian
fits, which is consistent with previous findings \citep{mor96, lei99,
ver01}.  The \ion{C}{4} equivalent widths are also listed in Table
~\ref{tbl-2}, along with the rest-frame continuum luminosity measured
at 1450\AA\ [$L_{\nu}(1450)$, expressed in ergs s$^{-1}$ Hz$^{-1}$].
The EW measurements may have errors resulting from the choice of
continuum placement; experiments with alternative continuum fits
suggest a systematic uncertainty of $\sim 15\%$.

Another parameter recorded in Table ~\ref{tbl-2} is the spectral index
$\alpha$ of the power-law fit (defined by $F_{\nu} \propto
{\nu}^{\alpha}$) that best approximates the continuum shape of each
individual object spectrum in the rest-frame.  In determining the
continuum solution, we tried to make use of the wavelength ranges
which contain pure continuum emission.  A reliable fit is obtained
when a wide separation of the continuum windows is available (e.g.,
$\sim$ 1100 -- 4000\AA), such that the Fe emission, which is strong
and prevalent in the UV and the blue part of the spectra of NLS1s, is
avoided.  In the present sample, there is only a small number of
objects for which the available spectra cover completely and/or extend
redward of the broad \ion{Fe}{2} and \ion{Fe}{3} emission-line
complexes.  For the sources spanning only the UV range, the measured
spectral index should be considered a lower limit, as the apparent
continuum can be strongly contaminated and reddened by the Fe
emission.  The full range of slope values and their distribution among
the sample are displayed in Figure ~\ref{Fig1}, along with the
calculated median and average values.  In identifying the Fe features,
and consequently the emissionless continuum windows, we used
information based on high resolution spectra of I Zw1, as provided by
\citet{ves01} in the UV bandpass, and by \citet{oke79} in the regions
redward of \ion{Mg}{2} $\lambda$2797.9.  The windows chosen in
determining the power law continuum shape are placed around 1140,
1285, 1320, 1350, 1450, 3810, 3910, 4040, 4150, 5470, 5770-5800, and
6210\AA, with the interval lengths ranging from 10 to 20\AA.  These
wavelength ranges are also used in fitting the composite spectra (see
Section~\ref{compo}), and they are illustrated in Figure ~\ref{Fig2}.

\section{The NLS1 composite spectra} 

In this section we present NLS1 composite spectra, and resulting
measurements of continuum and emission-line characteristics.  The use
of composite spectra is complementary to analysis of individual
spectra.  Composites offer higher S/N ratios, allowing
measurement of weak features, while also directly providing a
description of typical NLS1 spectral properties.  We examine the
results in relation to similar observations of broader AGN samples, in
order to explore the extent to which NLS1s represent a distinct
subclass.  These comparisons potentially provide useful tests for
discriminating between model scenarios for the NLS1 phenomenon.

\subsection{Overall continuum} \label{compo}

\begin{figure}
\figurenum{1}
\plotone{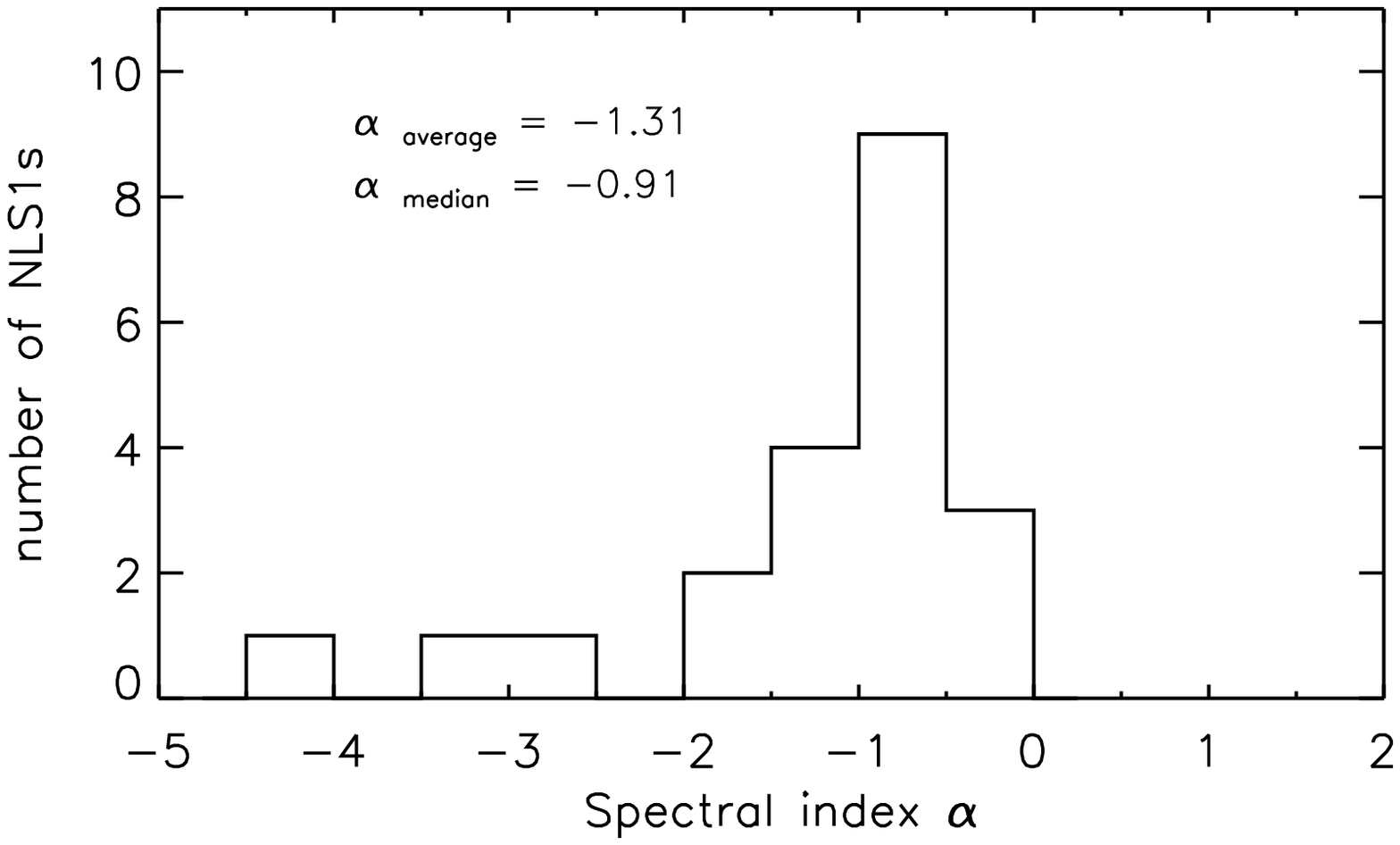}
\caption{Histogram of continuum spectral indices, $\alpha$, where
$F_{\nu} \propto {\nu}^{\alpha}$, for all but one (Mrk110) NLS1 in
the sample.  The average and median values are indicated.
\label{Fig1}}
\end{figure}

When building spectral composites, one of the most challenging tasks
is the generation of an underlying continuum that reflects the typical
appearance of the sample as a whole.  \citet{fra91} noted that ``there
is no `correct' way of co-adding spectra that exhibit differences on
many different scales.''  Our sample consists only of NLS1 sources,
which, as a subcategory of the AGN classification, share a set of
common attributes.  The spectra nonetheless show a substantial
diversity in their properties.  As Figure~\ref{Fig1} and
Table~\ref{tbl-2} reveal, the continuum shape, as described by the
power-law index $\alpha$, exhibits a considerable variety among
individual NLS1 spectra.  Because only very few objects in the sample
span the whole wavelength range, a simple median or average composite
displays artificial discontinuities at locations where the number of
contributing spectra changes significantly.  Some care is thus
required in order to use these composites to describe typical
continuum properties for NLS1 galaxies.

\begin{figure}
\figurenum{2}
\plotone{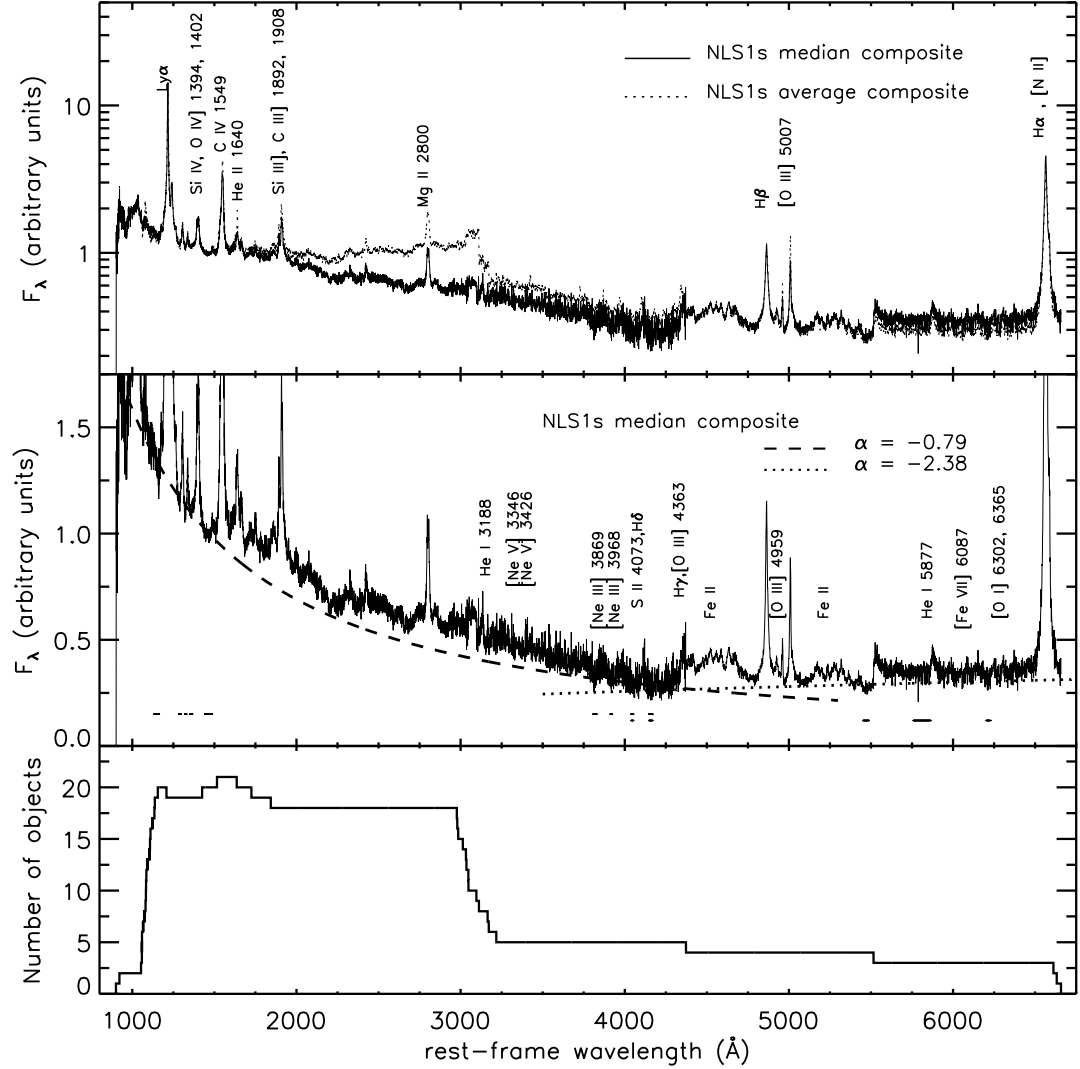}
\caption{{\it Top panel}: NLS1 composite spectrum plotted as
log(F$_{\lambda}$) vs. rest-frame wavelength, with the principal
emission features identified.  The flux has been normalized to unit
mean flux over the wavelength range 1430 \AA -- 1470 \AA.  {\it Middle
panel}: The median composite spectrum, plotted on a linear scale, and
zoomed near the continuum level for a better visualization of the weak
features in the optical range.  A more detailed UV line identification
is presented in Figure ~\ref{Fig4}.  The power-law continuum fits are
overplotted as dashed and dotted lines. {\it Bottom panel}: Number of
NLS1s contributing to the composite as a function of rest-frame
wavelength.
\label{Fig2}}
\end{figure}

Figure~\ref{Fig2} presents the NLS1 average and median composites,
constructed using data from individual objects normalized to the mean
flux in the wavelength range 1430 \AA -- 1470 \AA.  Each spectrum was
given equal weight, thus avoiding biasing the resulting composites
toward the brightest objects (i.e., those with the highest S/N).  The
median displays a smoother continuum than does the average, and is
well described by a single power law ($\alpha = -0.798 \pm 0.007$)
from just redward of Ly$\alpha$ to H$\gamma$.  Owing to the relatively
high S/N of the spectrum and the wide separation of the fitted
regions, the statistical uncertainty in the spectral index, as given
by a Chi-square minimization method, is quite small.  However, the
value itself is rather sensitive to the precise wavelength sections
employed in the fitting.  Redward of H$\gamma$, the continuum flux
density rises above the level predicted by the UV power-law, and is
best approximated by a separate power-law, with $\alpha = -2.38 \pm
0.01$.  Since these objects are low-luminosity AGNs, contamination by
the host galaxy starlight may contribute to this change in the
spectral shape.  Both fits are shown in the middle panel of Figure
~\ref{Fig2}; the continuum windows used in the power-law fits are
indicated as horizontal lines below the composite.  Prominent emission
features in the optical range are labeled.  The number of objects
contributing to the composites at each wavelength is presented in the
bottom panel of Figure~\ref{Fig2}.

The UV-blue continuum spectral index for this {\it HST} NLS1 composite
falls among the steepest values found in other AGN composite
measurements ($-1 < \alpha < -0.4$; see for example, Table 5 in Vanden
Berk et al. 2001)\footnote{\citet{rod00} have also reported evidence
that NLS1s are redder than other Seyfert 1s at optical wavelengths.}.
Such red continua were measured only in the \citet{zhe97} and
\citet{tel02} {\it HST} composite spectra ($\alpha = -0.99$, and
$\alpha = -0.71$ respectively), and in the \citet{sch01} sample of
{\it Sloan Digital Sky Survey} very high redshift quasars (average
$\alpha = -0.93$).  There are several possible reasons why these
particular samples show such steep continua, and we explore them in
the following paragraphs, in an effort to understand the origins of
the extreme continuum properties of the NLS1 composites.

One effect that may contribute to the soft continuum slopes of the
high $z$ QSOs and the {\it HST} sources derives from the restricted
wavelength baseline typically used in fitting the continua.  The
spectra of objects comprising these steep continuum samples make
available only short regions redward of Ly$\alpha$, in which the iron
contamination may produce an artificial rise in the continuum profile
\citep{tel02} .  However, in the NLS1 composites, the wide baseline
used in fitting the continuum permits a successful accounting for the
\ion{Fe}{2} and \ion{Fe}{3} complexes, therefore making the Fe
contamination an unlikely cause for their overall red
continua\footnote{The Fe contamination is the primary factor that
accounts for the difference between the spectral index that best fits
the NLS1 median composite and the median (and average) value of the
power-law index distribution of the individual continuum fits (see
Figure~\ref{Fig1}).  The majority of the objects in the NLS1 sample
cover only the UV wavelength region, where the power-law solution is
unable to properly estimate the true underlying level below the
\ion{Fe}{2} and \ion{Fe}{3} emissions, and as a consequence, steeper
indices are measured.}.

The steepness of the {\it HST} composites can also potentially be
attributed to an evolution of QSOs to softer spectra at lower
redshifts and/or an observational bias toward detecting sources with
harder continua at higher redshift \citep{fra93}.  The space-based
observations include a significant number of low redshift sources, and
this is the case for the NLS1 galaxies as well.  The same evolutionary
effect and the potential detection biases can also be responsible, at
least partially, for the redness of the NLS1 composites, especially
when compared with the ground-based composite spectra.

\begin{figure}
\figurenum{3}
\plotone{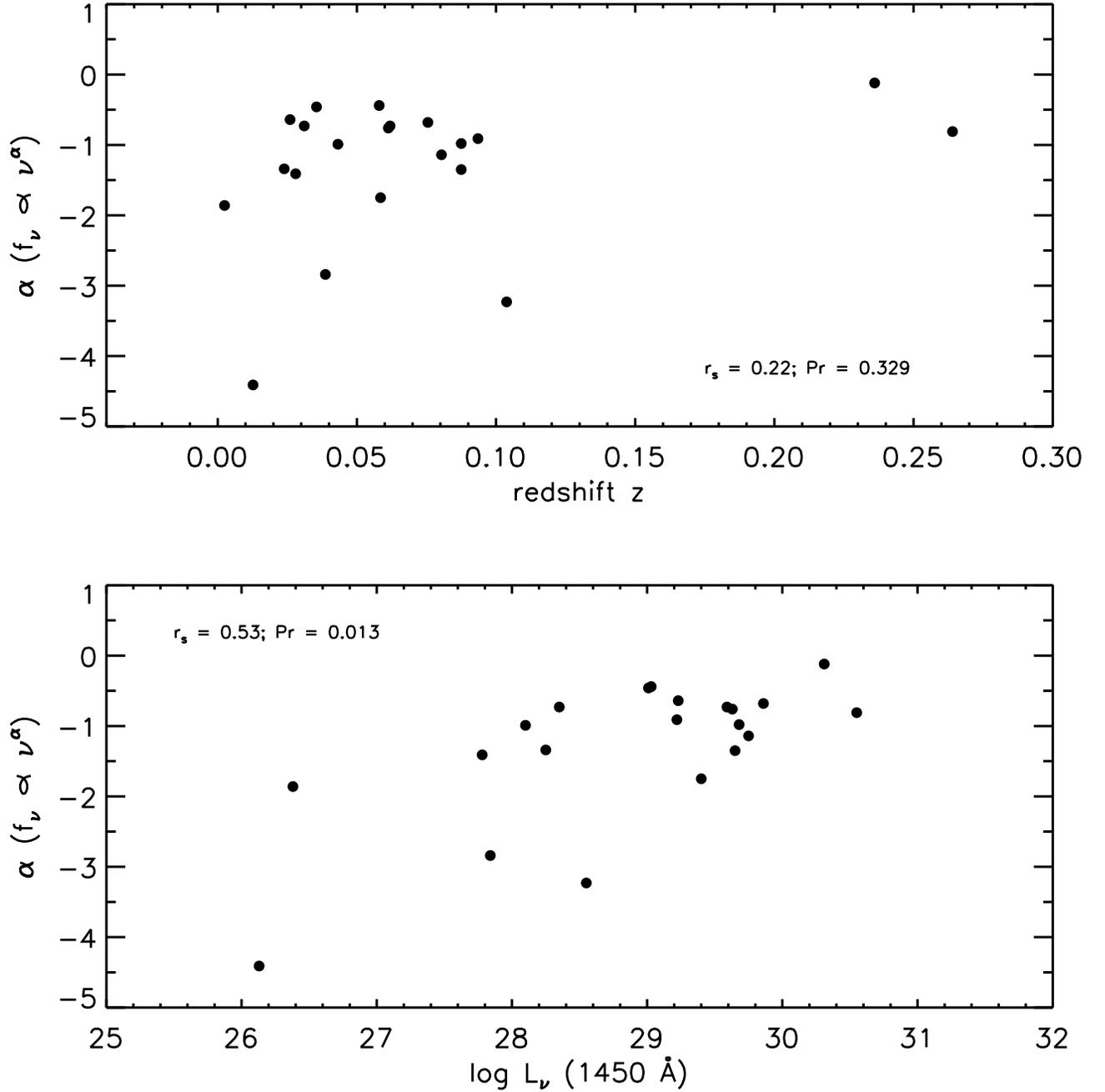}
\caption{Spectral indices plotted vs. redshift ({\it upper panel}) and
1450\AA\ luminosity ({\it lower panel}), with $L_{\nu}$ expressed in
ergs s$^{-1}$ Hz$^{-1}$.  The Spearman rank coefficient and the
probability of the correlation happening by chance are indicated.  The
error bars in both directions are smaller than the symbol size, and
therefore not indicated.
\label{Fig3}}
\end{figure}

The present sample shows little evidence of a correlation between
continuum slope and $z$, but a considerably stronger trend relating
$\alpha$ and luminosity (Fig.~\ref{Fig3}).  Regardless of its origin,
the luminosity dependence of spectral slope is probably largely
responsible for the relatively steep $\alpha$ for the NLS1 composite,
since the typical luminosity of our NLS1s is lower than that of
sources employed in most other quasar composites.

\subsection{Reddening and Absorption}

The luminosity dependence of spectral slope could be intrinsic to the
accretion source, but independent evidence suggests that it is mostly
attributable to luminosity-dependent reddening.  Internal dust, if
present, is expected to be accompanied by gas producing observable
absorption lines.  In our sample, the signature of strong absorption
near the systemic velocity of the host galaxy is present in almost
half the objects (see Table ~\ref{tbl-2}).  Median values of the
spectral indices corresponding to subsamples of sources with and
without strong resonance line absorption are $-1.34$ and $-0.73$
respectively, consistent with steeper (redder) continua in the
absorbed NLS1s.  Median logarithmic luminosity values for the absorbed
and unabsorbed subsamples are 29.01 and 29.60 respectively (mean
values are $28.4 \pm 0.4$ and $29.3 \pm 0.2$ respectively), thus
directly linking the presence of absorption with luminosity.

The location and physical state of the absorbers in NLS1s may be
important for understanding these objects.  Intrinsic absorption
related to the central accretion source is now known to be present in
a significant fraction of Seyfert 1 nuclei \citep{cre99}.  NLS1s
exhibit similarities in emission properties to low-ionization broad
absorption-line QSOs (BALQSOs), which \citet{bor02} has interpreted as
evidence for high luminosities relative to the Eddington value for
both classes of object; in the BALQSOs, the absorption is clearly also
closely related to the accretion process.  In the NLS1s, the
intervening gas may be a warm (highly ionized) absorber that is
potentially dusty \citep{kom99}; alternatively, the absorbing matter
may be related to the ``lukewarm absorber'' identified in case studies
by \citet{kra00} and \citet{cre02}, which may reside on kpc scales.
The luminosity dependence of reddening and absorption for the NLS1s is
consistent with a scenario in which the absorbing matter covers a
larger solid angle as seen by the central source in lower luminosity
objects, a natural expectation for interstellar matter in a disk-like
geometry of luminosity-independent scale-height, but with a
sublimation radius for dust that scales with luminosity, similar to
the `receding torus' model \citep{law91}.  If the absorbing medium has
a flattened distribution aligned with the host galaxy disk, we might
then expect a correlation between the host inclination and the
spectral (UV) color, as reported for several Seyfert 1s by
\citet{cre01}.  Inclination values are available in the literature for
16 of our sample members; the data do not show a statistically
significant trend with $\alpha$, however.  In summary, the extent to
which the UV/optical absorbers in NLS1 nuclei are associated with the
central accretion structure versus the normal host galaxy interstellar
medium remains ambiguous.

The nature of the dusty absorber in these galaxies can be further
tested by analyzing their soft X-ray (0.1 -- 2.4 keV) characteristics.
A ``lukewarm'' (low ionization) constitution of the gas would suggest
strong absorption, that translates into flatter observed soft X-ray
spectra.  If a ``warm'' (high ionization) absorber is present, steeper
X-ray continua should be measured \citep{gru98}.  The {\it ROSAT}
spectral indices ($\Gamma \equiv 1 - \alpha$) are available for the
majority of the NLS1s employed in this study, and are recorded in
Table ~\ref{tbl-2}.  Splitting the sample into absorbed and unabsorbed
objects, as before, and calculating the median value of the X-ray
spectral slope for each subgroup, should provide us a rough criterion
for distinguishing between the two types of absorbers.  The resulting
values are $\Gamma_{\rm absorbed\ spectra} = 3.4$ and $\Gamma_{\rm
unabsorbed\ spectra} = 3.1$, nominally implicating the presence of
highly ionized absorbers (mean values are $4.2 \pm 0.8$ and $3.3 \pm
0.2$ respectively); however, the difference in the median values is
small, and it appears likely that the absorbing medium in these
objects is described by a range of properties.

\subsection{The UV spectrum}

\begin{figure}
\figurenum{4}
\plotone{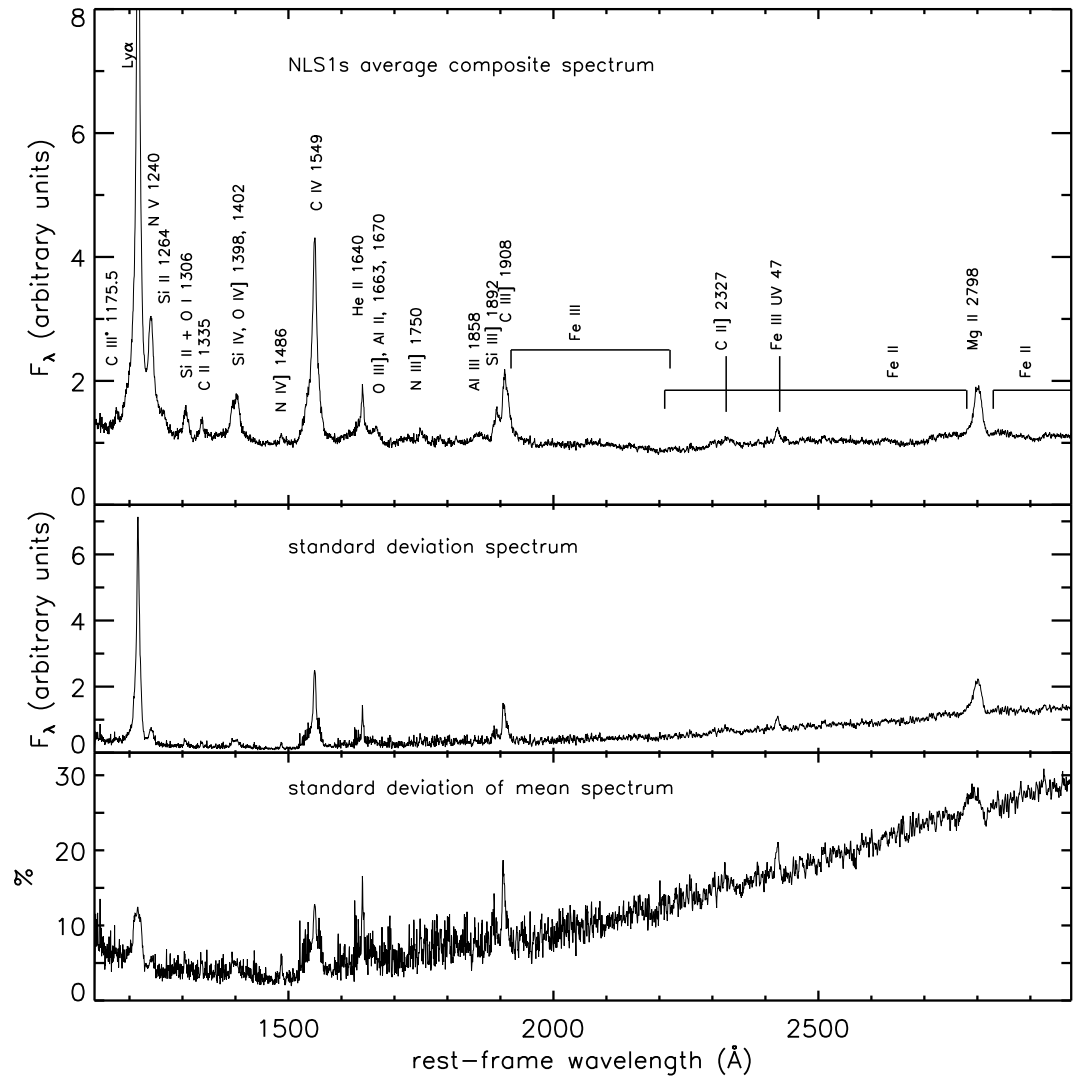}
\caption{{\it Top panel}: NLS1 average composite spectrum plotted for
the UV range only, where at least 18 objects are contributing.  The
same normalization as in Figure ~\ref{Fig2} is used.  The prominent
emission features are marked.  {\it Middle panel}: Flux standard
deviation in F$_{\lambda}$ relative to the average composite spectrum,
as a function of rest-frame wavelength for the individual spectra
comprising the composite.  {\it Bottom panel}: The standard deviation
of the mean, expressed in percent.
\label{Fig4}}
\end{figure}

Figure ~\ref{Fig4} shows the average spectrum in the UV range, where
most of the objects ($\geq 18$) in the sample are contributing.  The
standard deviation (RMS) and the standard deviation of the mean
spectra show the degree of variation between the members of the sample
and the uncertainty of the average composite as a function of
wavelength.  Near the wavelength corresponding to the common continuum
normalization ($\lambda$ $\sim 1450$\AA), the spectral variation
within the NLS1 sample is dominated by the modulations in the cores of
the strongest emission lines (Ly$\alpha$, \ion{C}{4}, \ion{He}{2}, and
\ion{C}{3}]).  In this respect, NLS1s behave like the other more
general samples of AGNs \citep{fra92,bro01}.  Away from the
normalization wavelength interval, the bulk of the spectral variance
is accounted for by differences in the individual continuum shapes and
noise.

\subsection{The optical composite}

\begin{figure}
\figurenum{5}
\plotone{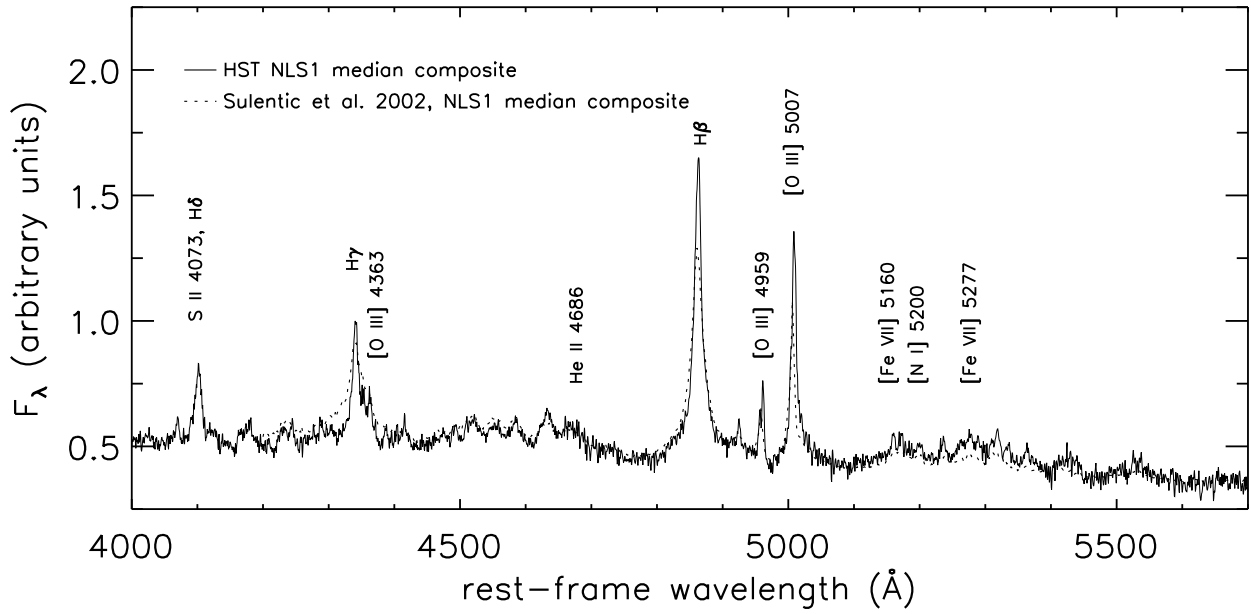}
\caption{ The optical range of the {\it HST} NLS1 median composite,
constructed using only the 3 objects that span the whole spectral
range.  The same flux normalization as in Figure ~\ref{Fig2} is used.
The \citet{sul02} NLS1 median composite, based on 24 ground-based
object spectra, is shown for comparison.  The strong similarity
between the two medians suggests that the {\it HST} NLS1 composites
are representative of these objects.
\label{Fig5}}
\end{figure}

For wavelengths $\lambda\ ^{>}_{\sim}\ 3000$\AA, the {\it HST} NLS1
spectrum is based on only a small number of objects.  Because fewer
than 5 NLS1s make up the composites at these wavelengths, the
individual source contributions are more pronounced, making it
difficult to identify and measure the spectral features (continuum and
emission lines) located at the transition points in the number of
contributing objects (e.g., at $\sim 4360$\AA, $\sim 5500$\AA).  A
smoother spectrum is constructed using only the three spectra that
span the whole wavelength range (Ark564, Mrk493, WPVS007), and is
presented in Figure ~\ref{Fig5}.  We compare our optical median
composite with the \citet{sul02} median NLS1 spectrum, constructed
from a much bigger sample (24 sources, ground-based spectra).  The two
spectra are very similar in both the continuum shape and the profile
and strength of emission features.  Small differences, like the
stronger peaks in the emission lines in the {\it HST} composite, are
mostly due to the different spectral resolution that characterizes the
two samples ( 4 -- 7\AA\ FWHM for the Sulentic spectra, $\sim$ 2\AA\
for the FOS-{\it HST} observations).  The good agreement between the
two medians suggests that the {\it HST} composites are representative
of typical NLS1 optical spectra, despite the fact that they are
constructed from such a small object sample.

\subsection{Emission Lines}  \label{lines}

The richness of the line emission in the AGN spectra is easily
distinguishable in the NLS1s due to their narrow widths.  Hence,
careful identification and intensive analysis of the NLS1 emission
lines have been performed, though to date, only on individual sources,
e.g. I Zw1 \citep[and references therein]{lao97, ves01}.  When
composites are built using relatively large samples, higher S/N is
achieved, and therefore, higher accuracy is expected in identifying
and measuring the emission lines.  We have been able to detect and
parametrize in the {\it HST} NLS1 composites a large number of
emission features.  Many are barely present or are heavily blended,
and therefore difficult to quantify in other AGN composites
\citep{fra91, van01, zhe97, tel02}.

The emission lines are measured from the average composite, as it
presents a higher S/N ratio than the median spectrum.  An
exception is made for the wavelength interval $\sim 2900 - 3170$ \AA,
where the emission features in the average spectrum are distorted by
the jumps present in the continuum shape (see Section~\ref{compo}),
and where the median is used.  Line fluxes, strengths (EWs), widths
(FWHM) and line shifts relative to the laboratory central wavelengths
($\Delta$v) are listed for all detected features, along with their 1
$\sigma$ error bars, in Table~\ref{tbl-3}.  The lines have been
identified by matching wavelength positions and respective relative
strengths of features found in the \citet{fra91}, \citet{zhe97},
\citet{van01} and \citet{tel02} composites, and in the detailed
analyses of I Zw1 \citep{lao97, ves01}.  Tentative identification of
newly discovered features is based mainly on data available from the
Atomic Line List\footnote{The Atomic Line List is hosted by the
Department of Physics and Astronomy at the University of Kentucky;
http://www.pa.uky.edu/$\sim$peter/atomic.} and \citet{ver96}.

All measurements, including the peak positions of each emission line,
$\lambda_{mean\_rest}$, and the respective errors are generated using
the task $specfit$ \citep{kri94} as implemented in the
IRAF\footnote{The Image Reduction and Analysis Facility (IRAF) is
distributed by the National Optical Astronomy Observatories, which is
operated by the Association of Universities for Research in Astronomy
Inc.  (AURA), under cooperative agreement with the National Science
Foundation.}  software package.  The method employs line and continuum
spectral fitting via an interactive ${\chi}^2$ minimization.  The fit
is performed in 14 separate spectral intervals; their lengths varied,
in order to keep a reasonable number of parameters, from $\sim$
150\AA\ in the UV-blue region, rich in prominent emission features, to
$\sim$ 500 -- 900\AA\ in the optical range.  The equivalent widths are
measured relative to the resulting local continua, which differ from
the shape estimated in Section ~\ref{compo}, mainly due to the fact
that the average and not the median composite is used in the fitting
process.  Comparisons indicate that line fluxes measured in the
average and median composites are consistent to within $\la 10\%$ in
most cases.

For most features, the line emission is modeled using single
Lorentzian profiles.  However, the most prominent lines required two
components, a narrow and a broad Lorentzian, the broad one being in
general blueshifted relative to the narrow component.  The broad and
prominent Fe emission surrounding and redward of
\ion{Mg}{2}$\lambda2798$ is not measured; thus, line measurements for
2000 \AA\ $\la \lambda\ \la$ 4250\AA\ may include some Fe
contamination, and therefore, a high level of uncertainty.  For the
optical wavelength range, redward of H$\gamma$, we use in the fitting
process the empirical \ion{Fe}{2} template obtained by \citet{bor92}
from the I Zw1 spectrum.  No broadening of the iron template was
necessary in order to match the line width of the NLS1 composite
spectra.  In fitting forbidden line doublets, the line pairs are
assigned to have common velocity widths and offsets, and their flux
ratios are constrained to values determined by branching ratios, when
appropriate.

\begin{figure}
\figurenum{6}
\plotone{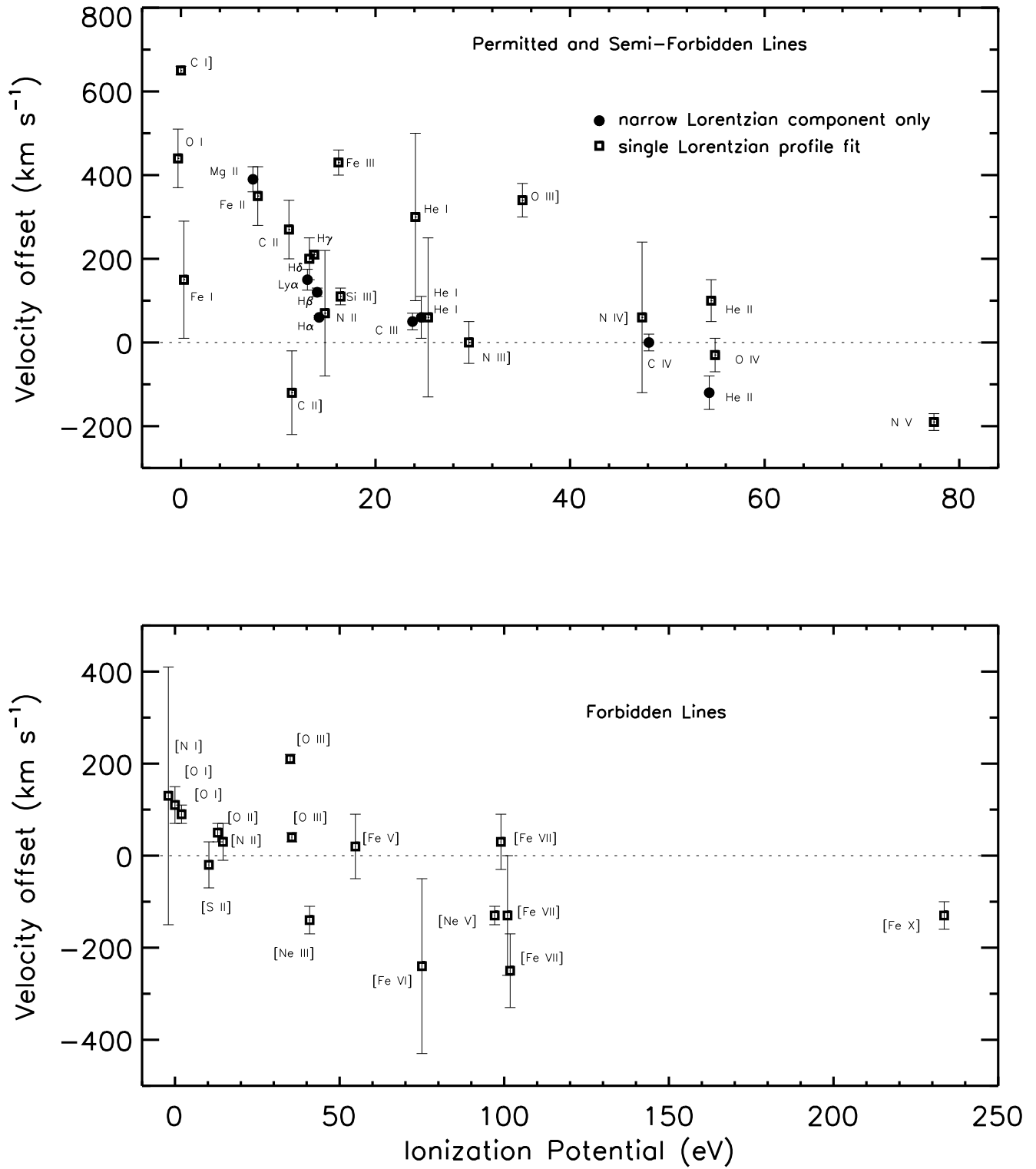}
\caption{ Emission-line velocity offsets, relative to the rest frame
(defined by \ion{C}{4}$\lambda1549$), as a function of ionization
potential for selected emission lines (see text).  Error bars show the
1 $\sigma$ uncertainty in the velocity measurement.  Permitted and
semi-forbidden lines are shown in the {\it top panel}, and forbidden
lines are shown in the {\it bottom panel}.  Only one measurement is
shown for each line, and the point type indicates the adopted
profile. Overlapping points are slightly offset horizontally from each
other for clarity. The points are labeled by ion.
\label{Fig6}}
\end{figure}

Because the composites were constructed using redshifts based on the
position of a single emission-line, \ion{C}{4} (see
Section~\ref{data}), a check for systematic velocity offsets for other
lines can be performed.  Such line shifts, which are definitely
present for many of the lines listed in Table ~\ref{tbl-3}, have been
detected previously in other AGN samples \citep{gas82, esp89, tyt92,
mci99, van01}.  Note that most of the lines are redshifted relative to
\ion{C}{4}, the reference line, and that the largest values of the
recorded velocity shifts correspond to the lowest ionization states
(e.g., \ion{Si}{2}, \ion{C}{2}, \ion{O}{1}).  Figure ~\ref{Fig6} shows
the velocity shifts versus the ionization potentials, compiled for all
emission lines that are stronger than \ion{C}{2}$\lambda1335$, i.e.,
EW $\ga 4.0$ \AA, and that have well defined non-blended peaks.  We
treat the permitted and the forbidden lines separately since they may
have very different origins.  The values for the ionization potentials
are chosen such that, for the recombination lines of H and He, they
express the energy necessary to ionize the respective state for later
recombination, while for the collisionally excited lines from heavy
elements, they represent the energy that is needed to create the
ionization state.  For the most prominent features, which are fitted
with two components, we consider only the velocity shift measurements
given by their narrow components (they are identified by different
symbols in Figure ~\ref{Fig6}, upper panel); this treatment is
justified by the fact that the reference wavelength is chosen based on
the location of the peak of \ion{C}{4}, which is measured using the
narrow component fit\footnote{Single Lorentzian fits for these
features result in line shifts that differ by $^<_{\sim} 100$ km
s$^{-1}$ from the ones given by the narrow component only; this is due
to the fact that the cores of these relatively narrow lines account
for the bulk of the flux.}.  This approach is consistent with
measurements of velocity offsets performed in previous similar studies
of other AGN samples (e.g. Vanden Berk et al. 2001).

The results presented in Figure ~\ref{Fig6} provide evidence for an
anticorrelation between the velocity shifts and the degree of
ionization of the emitting species.  The trend is definitely present
in both categories of emission features.  For the forbidden lines,
significant velocity shifts, usually blueshifts relative to the
systemic velocity, have been detected in the past for the high
ionization transitions \citep{pen84, app91}; however, the evidence for
a correlation similar to that presented in this study has only
recently been revealed by measurements from the high S/N {\it Sloan
Digital Sky Survey} quasar composite spectrum \citep{van01}.  In both
types of lines measured in the NLS1 sources, the amplitude of this
correlation appears similar to that present in other, more
heterogeneous Sy1/quasar samples\footnote{The velocity shifts may be
unusually large in I Zw 1; see \citet{lao00}.}.  Although the
ionization potential seems to govern the magnitudes of the velocity
offsets, there is a significant amount of scatter in this empirical
relationship, in both Sy1/quasars and NLS1 objects, suggesting the
influence of other parameters as well.  The scatter in the NLS1 trend
may be amplified by wavelength zero-point calibration uncertainties,
particularly for the FOS spectra.

The origin of the velocity offsets between the AGN emission lines is
not well understood, but a promising means of interpreting this
behavior is in terms of the disk-wind model for the broad-line region
(BLR).  In this picture, the high ionization lines are produced in
outflowing winds, accelerated by radiative line driving, that arise
from the accretion disk, which is the base for the low-ionization
emission \citep{mur98, pro00, lei01}.  Additionally, the disk and/or
the radiative outflow itself, believed to proceed from near the plane
of the disk, is assumed to be optically thick, so the emission from
the receding wind is obscured.

The fact that NLS1s and other broad-line AGNs show similar line shifts
provides additional constraints for the physical models that best
characterize the NLS1s.  The preferred explanations for the primary
drivers of the NLS1's extreme measured properties postulate either
higher ratios of their luminosity to the Eddington luminosity
\citep{pou95}, or a more pole-on view of an assumed disk-shaped BLR
\citep{bra00}.  The velocity offsets between lines reported here, and
their similarity to those found in less restricted AGN samples, may
present a challenge to either NLS1 scenario, if the velocity
differences stem from a disk-wind phenomenon or something similar.
For the orientation model, it would be surprising if the typical
line-of-sight {\sl differences} in line velocity remain essentially
unchanged when the orientation is such that the line {\sl widths} are
diminished.  For the high Eddington ratio picture, the offsets
ultimately trace the disk outflow velocity, which is expected to scale
with the local orbital speed, hence large velocity differences
combined with small widths are again unexpected.  The solution in
either case may ultimately involve the fact that the narrow line
widths characterize only a subset of emission features in NLS1s,
notably including the H$\beta$ feature, while other lines
(particularly in the UV) still reveal high-velocity gas components.
Indeed, as discussed by \citet{wil00}, and elaborated by
\citet{sha03}, the UV line widths do not correlate in a simple way
with H${\beta}$ width or with other NLS1 defining properties.

\section{Are NLS1s the analogues of high $z$ QSOs?}

Based on the apparent similarity of some of the emission properties of
NLS1s and $z \ga 4$ QSOs, \citet{mat00} suggested a connection between
these two classes of AGNs, in the sense that the NLS1s are the low
redshift, low luminosity analogues of the high $z$ quasars.  In the
proposed picture, both categories of sources are in an early
evolutionary phase, in which accretion proceeds at or near the
Eddington limit.  This scenario has several appealing aspects for
explaining NLS1 phenomena, but additional tests are desirable to
verify this idea, and especially to gauge its applicability to the
high $z$ sources.

One of the key arguments employed in support of a NLS1 -- high $z$ QSO
association was the initial report of narrow UV lines in $z\
^{>}_{\sim}\ 4$ quasars by \citet{shi97}.  While the presence of an
enhanced low-velocity component in the high redshift quasars was later
confirmed by \citet{con02}, the UV line profiles do not necessarily
provide an adequate basis for linking the high $z$ QSOs to a NLS1
classification, as this is based on optical lines.  In fact, as noted
above (Section ~\ref{lines}), the UV and optical emission-line
properties are substantially independent \citep{sha03}.

Another possible similarity between high $z$ quasars and NLS1 galaxies
is a high metallicity ($Z$) in their emitting gas.  In quasars at $z
\ga 3$, super-solar heavy element enrichments have been derived in
specific abundance studies based on both emission \citep{die99, die00,
die03} and absorption properties \citep{ham97}.  The high gas-phase
metallicity is believed to derive from a recent episode of vigorous
star formation.  For the NLS1s, several lines of argument have
suggested the presence of enhanced metallicity.  One of the potential
indicators of high abundances in these objects is their
characteristically strong \ion{Fe}{2} emission \citep{col00}.
However, the excitation of this emission is complicated and its
strength does not necessarily map in a simple way to the Fe abundance
\citep{ver99}.  Moreover, if the Fe is largely produced in Type-Ia
supernovae, as would be expected, the evolution of the progenitor
population and the resulting enrichment should occur on a timescale
$\ga$ 1Gyr; consequently, a high iron enrichment is not a strong
indicator of youth, and therefore, would not support the hypothesis
of NLS1s being galaxies in the making.

A better tracer of abundances in NLS1 galaxies is provided by
nitrogen.  This element is of particular interest since its production
is believed to be dominated by secondary enrichment, which translates
into N/H $\propto$ $Z^2$.  \citet{nag02} have recently discussed the
forbidden line spectra of NLS1s, including [\ion{N}{2}] $\lambda$6583,
and argued that the observed line ratios suggest higher metallicities
than in the Sy1s; as the authors also point out, however, this
statement is very model-dependent.  More robust diagnostics are
obtained from studies of the \ion{N}{5} $\lambda1240$ feature, and in
particular, measurements of its strength relative to \ion{C}{4}
$\lambda 1549$ and \ion{He}{2} $\lambda 1640$ (Hamann et al. 2002, and
references therein).  Evidence for systematic trends of stronger
\ion{N}{5} and weaker \ion{C}{4} with increasing EV1 (which seems to
trace NLS1 behavior, see Section 1) has been reported by
\citet{wil99}.  Based on the same flux ratios, \citet{she02} showed
that the NLS1 sources deviate significantly from the well known
relationship between $Z$ and luminosity ($L$) in AGNs
\citep{ham93}, by exhibiting higher $Z$ at a given $L$.
Interestingly, several of the nine extreme NLS1s used in their study
have line ratios as high as those measured in the most luminous high
$z$ QSOs.  Measurements for larger samples of NLS1s are clearly
desirable to verify these findings.

In this section, we investigate the NLS1 -- $z\ ^{>}_{\sim}\ 4$ QSO
analogy by directly comparing their emission-line properties.  As most
of the line emission studies of the NLS1 objects are in the optical
regions, and the corresponding (rest-frame) data for $z ^{>}_{\sim}\
4$ are almost nonexistent, the UV spectral observations are the only
accessible tool to exploit in such a comparison.  We have available
both NLS1s and high $z$ quasar UV data, for samples that permit an
adequate statistical analysis of their emission characteristics.  Our
study employs both comparisons of their composite spectra, and an
eigenvector analysis that determines the degree to which the spectral
variances throughout the samples share common properties.  The NLS1
objects and composites used here are described in Sections ~\ref{data}
and ~\ref{compo}.  For the $z\ ^{>}_{\sim}\ 4$ QSO sample, we made use
of 44 spectra of non-BAL quasars, presented by \citet{con02} and
\citet{sch91}, which span the 1100 -- 1700\AA{} rest-frame range.

\subsection{Direct comparison of composite spectra}

\begin{figure}
\figurenum{7}
\plotone{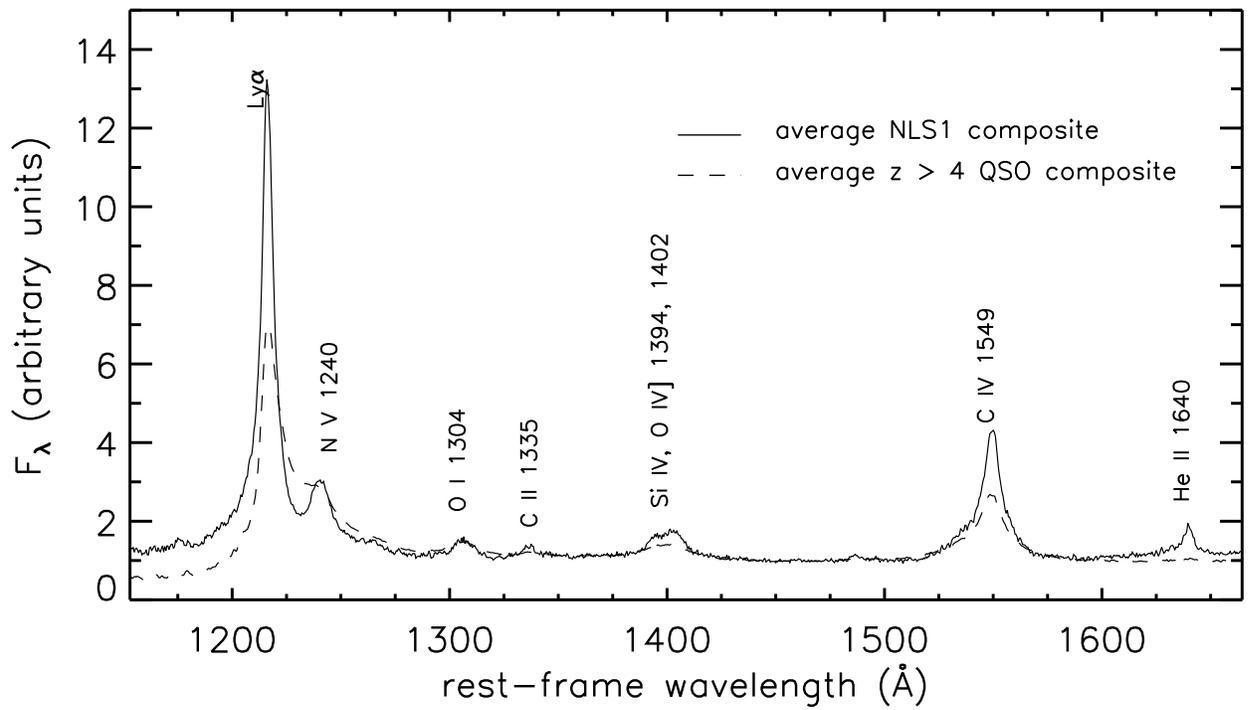}
\caption{ Direct comparison between NLS1 and $z\ ^{>}_{\sim}\ 4$ QSO
average composite spectra.  The same normalization as in Figure
~\ref{Fig2} is used.
\label{Fig7}}
\end{figure}

Figure~\ref{Fig7} shows the NLS1 average composite
(Section~\ref{compo}) overplotted on the $z\ ^{>}_{\sim}\ 4$ QSO
average composite \citep{con02}.  In a first approximation, the
spectra agree well: similar continuum shape and the same strong
emission lines, with comparable profiles.  The pronounced Ly$\alpha$
forest in the $z \ga 4$ composites does not reflect an intrinsic
difference in the nature of the emission sources, but the expected
increase in the opacity of the intergalactic medium at larger
redshifts.  

\begin{figure}
\figurenum{8}
\plotone{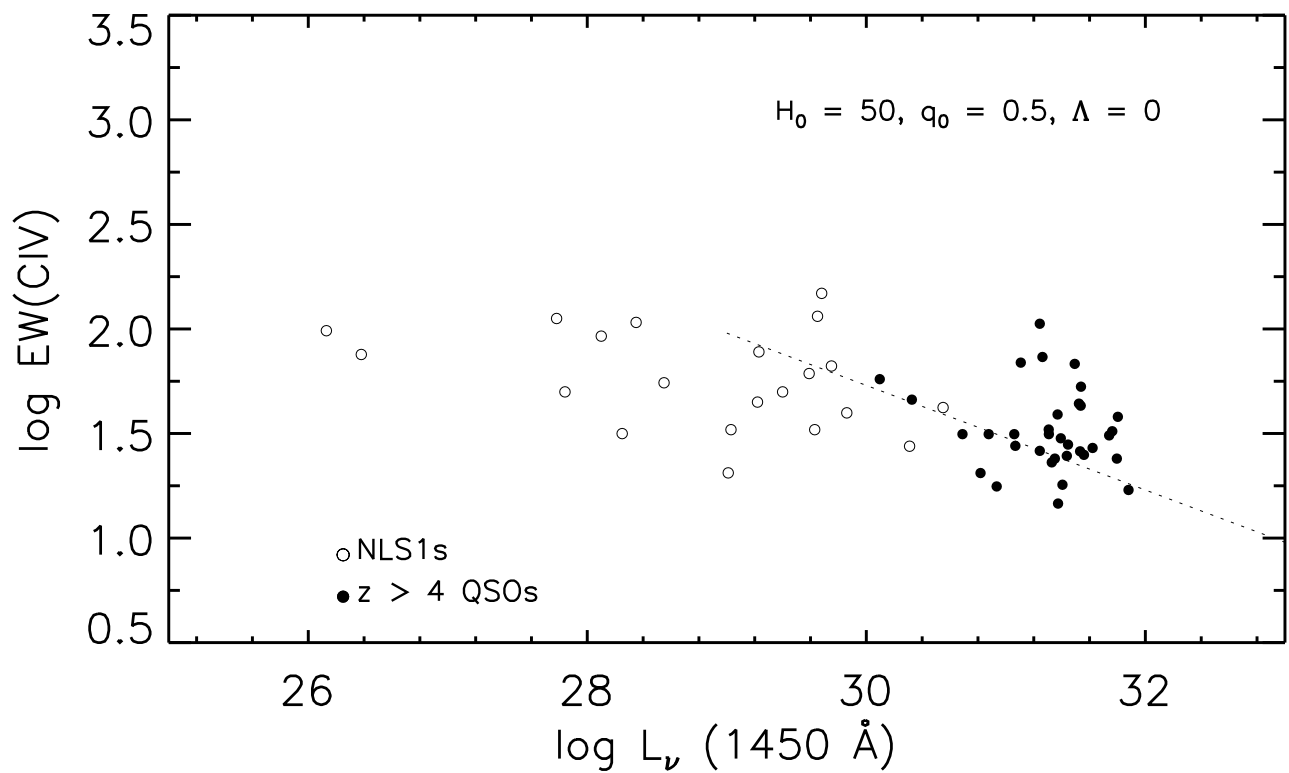}
\caption{ Rest-frame EW of the \ion{C}{4} emission line, in \AA, as a
function of 1450\AA\ luminosity, in ergs s$^{-1}$ Hz$^{-1}$, for
individual NLS1 and $z\ ^{>}_{\sim}\ 4$ QSO spectra. The Baldwin
relation found by \citet{osm94} for a sample of 186 luminous quasars
[log$L_{\nu}(1450)\ ^{>}_{\sim}\ 29$, $0 < z < 3.8$] is shown as a
dotted line.
\label{Fig8}}
\end{figure}

The most evident distinction between the NLS1 and high $z$ quasar
composite spectra is in the strength of the principal emission
features: Ly$\alpha$, \ion{Si}{4}+\ion{O}{4}], \ion{C}{4} and
\ion{He}{2}.  This difference in the line strengths is a clear
manifestation of the Baldwin Effect (the inverse equivalent width --
luminosity relationship, Baldwin 1977).  The two samples differ
significantly in their average luminosity, with the lower luminosity
objects, the NLS1s, exhibiting stronger lines.  Figure~\ref{Fig8}
shows the individual measurements of the rest-frame continuum
luminosity, L$_{\nu}$(1450\AA), and the strength (EW) of the
\ion{C}{4} line, plotted for both the NLS1s and high $z$ quasars.  The
Baldwin Effect correlation in \ion{C}{4} is clearly present over the
combined sample; the segregation in luminosity of the two samples is
also evident \footnote{The flattening of the Baldwin correlation at
low luminosities has been reported in several previous studies; see
\citet{osm99} for discussion and references.}.  Besides the
basic Baldwin trend, the comparison in Figure~\ref{Fig7} shows also
the known trend with ionization: steeper Baldwin Effect for more
highly ionized species \citep{die02}.  The difference in the strength
of the emission lines in the NLS1 and high $z$ QSO composites is
likewise most obvious in Ly$\alpha$ and the high ionization features
(\ion{Si}{4}+\ion{O}{4}], \ion{C}{4} and \ion{He}{2}), and almost
absent in the low ionization lines (\ion{O}{1}, \ion{C}{2}).  As found
in previous studies of the Baldwin relationships, the
\ion{N}{5}$\lambda$1240 emission proves to be the exception for which
the line strength remains nearly independent of luminosity.

Another important issue that can be examined via comparison of the
composite spectra is the degree of similarity in the chemical
enrichment for the two source types.  As noted before, the broad line
region enrichments in both $z\ ^{>}_{\sim}\ 4$ QSOs and NLS1s have
been evaluated based on the \ion{N}{5}/\ion{C}{4} and
\ion{N}{5}/\ion{He}{2} line ratios, as best indicators of the overall
metallicity.  Figure~\ref{Fig7} shows that the \ion{N}{5} emission
feature is equally strong in the NLS1 and high $z$ QSO average
composites.  Deblended measurements of this line obtained from
detailed fits of the
Ly$\alpha$+\ion{N}{5}+\ion{Si}{2}$^{\ast}$+\ion{Si}{2} emission-line
complex confirm this result.  \ion{C}{4} and \ion{He}{2} are however
much stronger in the NLS1 spectra than in those of the $z\
^{>}_{\sim}\ 4$ QSOs \footnote{The same results are obtained when
median composites are used for comparison.}.  These trends result in
lower \ion{N}{5}/\ion{C}{4} and \ion{N}{5}/\ion{He}{2} line ratios,
and presumably lower average metallicities, in the NLS1 sources than
in the high $z$ quasars, and thus do not support a strong connection
between the two classes of objects.  The results of this comparison
differ from those of \citet{she02} who argued that the line strengths
suggest comparable metallicity for the NLS1s and the high luminosity
QSOs.  However, their conclusion was based on a very small number of
NLS1s; also, the high abundances in these sources were supported by a
relatively large \ion{N}{5}/\ion{C}{4} ratio, but not clearly
confirmed by \ion{N}{5}/\ion{He}{2}.  The discrepancy between their
findings and ours may be related to the details of the samples used in
each case.  With the larger sample size employed here, we tentatively
conclude that the abundances in NLS1s and high $z$ QSOs are in fact
characteristically different; further study with larger sets of object
spectra would clearly be desirable.

\subsection{Principal Component Analysis}

Additional information on the NLS1 -- $z \ga 4$ QSO relationship can
be obtained by the means of a Principal Component Analysis (PCA), a
mathematical decomposition of a set of properties describing the
sample into a smaller number of eigenvectors that can account for the
bulk of the total variance present in the data.  Due to the fact that
quantitative measurements of the spectral properties are in general
subjective to the chosen parametrization, we chose to apply the PCA
method directly to the observed spectra \citep{fra92}.  We proceed
with this analysis for the two samples separately, and for the
combined set as well.  This approach should allow for further
identification of potential common behaviors and their distribution
among the two object classes.

\begin{figure}
\figurenum{9} 
\plotone{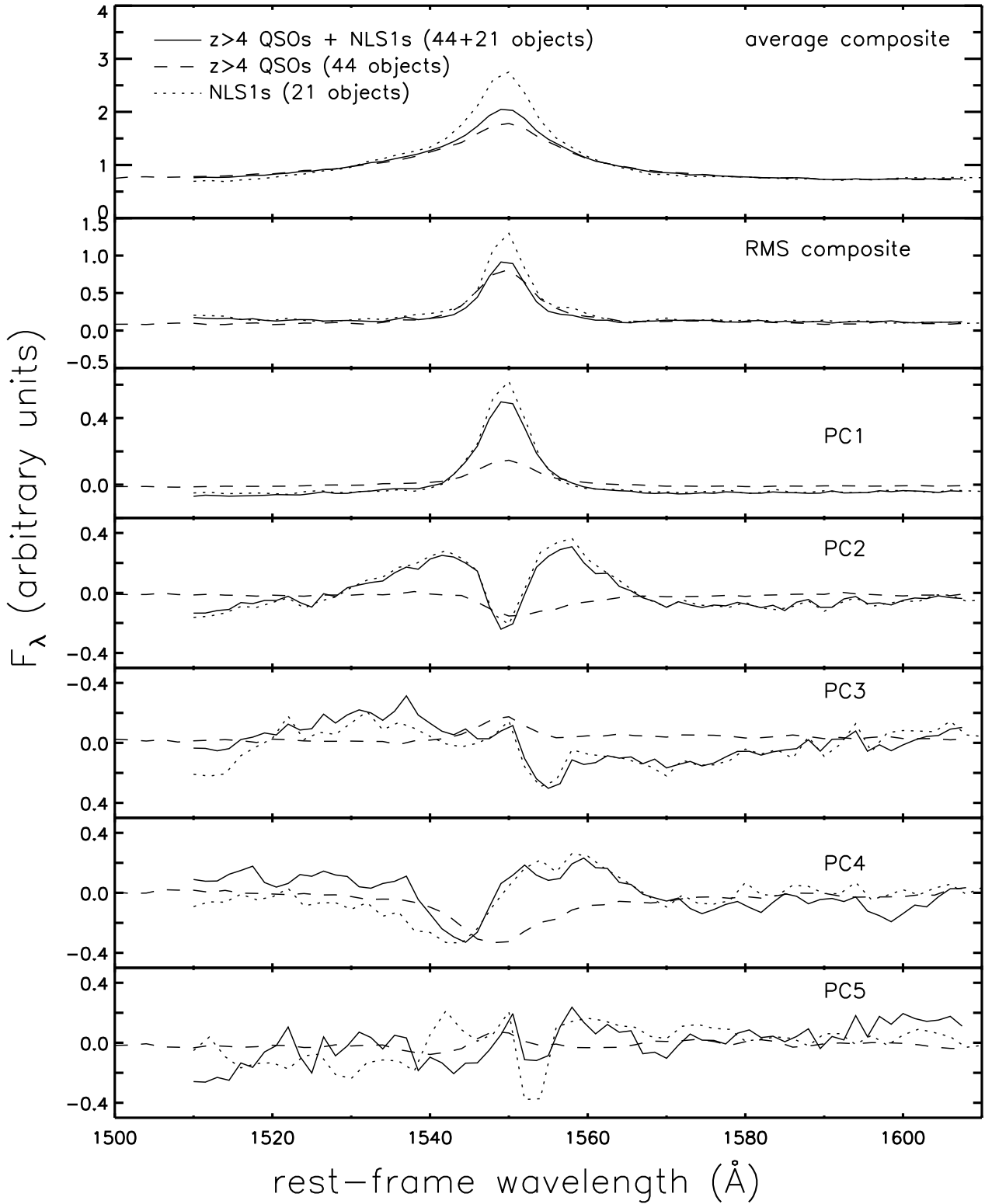}
\caption{ Mean and standard deviation spectrum and the first 5
principal components given by PCA, performed for the \ion{C}{4} region
only, applied to the NLS1 sample (dotted line), the $z\ ^{>}_{\sim}\
4$ QSOs (dashed line), and the combined sample (continuous line).  The
principal components show similar modulations in the spectral
variation in both categories of objects; the amplitude differs.
\label{Fig9}}
\end{figure}

For the NLS1 data, the spectral range with the most extensive coverage
among the sample objects (21 out of 22) spans the region around the
\ion{C}{4} line, and this bandpass is thus optimal for detailed 
statistical investigation. In this wavelength
range, the $z\ ^{>}_{\sim}\ 4$ QSO sample is well represented by all
44 object spectra. Therefore, we conduct the PC analysis on this
interval, where the highest S/N is expected.  Figure
~\ref{Fig9} shows the mean spectra, the RMS spectra, and the first
five principal components (PCs), ordered by the fraction of the sample
variance for which each accounts, for the NLS1s, high $z$ QSOs, and
the combined sample.  The $range$ of properties exhibited by these
objects is illustrated by the PCs and by their contribution to the
total spectral variation.  Principal Component 1 (PC1) is dominated by
line-core modulations (compared with the emission in the average
composites), and therefore, it can be considered primarily as a
measure of the strength of the lines.  This is a similar result to
what Francis et al. (1992) obtained with PC analysis for a larger and
more heterogeneous sample (232 objects), from the Large Bright Quasar
Survey (LBQS).  Aside from a scale factor, this first principal
component is very similar for the three samples.  Its profile differs
in the two individual samples (FWHM(NLS1s) = 2050 km s$^{-1}$, FWHM($z \
^{>}_{\sim}\ 4$ QSOs) = 3100 km s$^{-1}$); as might be expected, the
NLS1 core contribution is narrower than that in the high $z$ QSOs.
The second component, PC2, accounts for spectrum-to-spectrum
variations present in the wings of the lines.  This is again exhibited
by both samples but in a much larger proportion by the NLS1s.  Also,
the line-core and the line-wing modulations show opposite trends,
suggesting that objects with prominent wings possess weak cores.
Higher-order components continue to display, in different proportions,
the core and the wing variations, but their intricacies become
difficult to characterize and compare.

Associating physical explanations to the properties exhibited by PCs
is difficult, but additional information can be gained from the
statistics of the PCA results.  Table ~\ref{tbl-4} lists the
proportions (fractions of total variance) contributed by the first
five PCs for each individual sample and for the combined one.  The
numbers indicate that the NLS1 sample can be better represented than
the $z \ ^{>}_{\sim}\ 4$ quasar sample by a low number of principal
eigenvectors.  The difference in the PCs' individual and cumulative
proportions indicate that NLS1 sources comprise a more
spectroscopically compact sample than the high $z$ quasars.  Another
basis for evaluating these results is a statistical comparison of the
distributions of the weights of the first two principal components for
the NLS1 and $z \ga 4$ QSO objects.  In the PCA run for the combined
sample, i.e., for a common set of eigenvectors representing the NLS1s
and the high $z$ QSOs, the Kolmogorov-Smirnov (KS) test shows that the
possibility of the two data sets being drawn from the same parent
population is not excluded, but the evidence for this is weak: the KS
probabilities are 0.047 and 0.188 for weights of PC1 and PC2
respectively.

As a consistency test, we also performed this analysis for a much
larger spectral range, covering Ly$\alpha$ to \ion{He}{2} bandwidth,
but with a lower number of NLS1s (only 16 objects), and therefore
lower statistics; the main results remained unchanged.  NLS1s and
$z\ga 4$ QSOs are probably not close spectroscopic or physical
analogues.

\section{Conclusions}

In this paper we present an analysis of all publicly available spectra
for NLS1 galaxies in the {\sl HST} archive. The resulting sample of 22
NLS1s with spectra spanning the UV-blue wavelength range is the
largest that has been used, to date, in emission line studies of these
objects.  We employed these data to construct composite spectra
(average and median), and thus, to characterize in detail the typical
spectral properties of the NLS1 class.

The resulting composite spectra are used to estimate the strengths of
a large number of emission lines, and to quantify the continuum shape
over a broad bandpass.  Power-law fits to the continuum indicate a
discrepancy from the results obtained from more general AGN composite
studies: NLS1s have steeper UV-blue spectra.  Possible explanations
include intrinsic reddening and a trend of the low redshift sources to
have intrinsically softer continua.  A relation between the continuum
shape and the redshift is not readily evident in this sample.  A
significant correlation is however observed between the spectral slope
and luminosity, indicating that the redness of the NLS1s is related to
the low luminosity of these objects.  Moreover, the apparent
connection between the UV resonance absorption lines and luminosity
suggests that the steep slopes measured in these objects are due at
least partially to reddening.  The ionization state of the absorbing
material and its relationship to the accretion source are not well
determined, however.

The NLS1 composites additionally allow us to quantify emission-line
velocity offsets.  The correlation between the velocity shifts and the
degree of ionization that is found in normal broad-line AGNs is also
present in the NLS1 sources, in both permitted and forbidden lines.
This result may be of interest for comparison with NLS1 model predictions.

The NLS1 data permit further investigation into the proposed analogy
between these sources and the $z \ ^{>}_{\sim}\ 4$ quasars.  Previous
work suggested that both may be described by a high accretion rate and
super-solar metallicities.  The comparative study that we conduct
based on the NLS1 and high $z$ quasar average UV emission properties
reveals a significant contrast between their spectral characteristics.
The composite spectra exhibit primarily the anticipated differences
associated with the Baldwin Effect, i.e., relatively stronger lines in
the less luminous sources.  Nonetheless, the comparison makes it
evident that the metal enrichment of the surrounding gas in the high
$z$ QSOs is higher than that in the NLS1 objects.  This result does
not support the hypothesis that these two types of objects are similar
in their detailed physical characteristics or evolutionary phase.

Additional confirmation of the differences exhibited by the two types
of objects comes from a principal component decomposition applied
directly to the spectra.  The spectral PC analysis indicates that the
$z \ga 4$ sources are statistically likely to share only part of their
emission properties with the NLS1 galaxies; in particular, the
low-velocity component, although enhanced in the high $z$ QSOs, is
less prominent than in the NLS1 objects.  The statistics show that the
$z\ ^{>}_{\sim}\ 4$ QSO phenomenon is controlled by a broader, more
heterogeneous family of properties than those governing the NLS1s; the
NLS1 spectroscopic features can be reconstructed from a smaller number
of parameters (eigenvectors) than that necessary in the case of $z \ga
4$ sources.  In conclusion,  NLS1s and high $z$ quasars are spectroscopically
disparate, and it is therefore doubtful that a close physical connection
exists between these source types.

\acknowledgements

The authors thank Todd Boroson for providing the I Zw1 Fe template
spectrum in digital form, Don Schneider for the high $z$ QSO spectra
in digital form, Paul Francis for access to his PCA software, and Fred
Hamann for helpful discussions.  We also gratefully acknowledge Ari
Laor and the anonymous referee for providing comments on the original
manuscript that resulted in improvements.  Support for this research
was provided by NASA through grant GO-09143.02 from the Space
Telescope Science Institute, which is operated by the Association of
Universities for Research in Astronomy, Inc., under NASA contract
NAS5-26555.

\clearpage

\clearpage

\begin{deluxetable}{lrrrrcrrrr}
\tablecolumns{5} \tablewidth{0pt} \tablecaption{NLS1 Sample
\label{tbl-1}} \tablehead{ \colhead{Object} & \colhead{Instr.} &
\colhead{Gratings} & \colhead{Coverage\tablenotemark{a}} &
\colhead{Ref\tablenotemark{b}}} 
\startdata 
Ark 564 &FOS&G130H,G190H,G270H,G400H,G570H &1087--6817& 6,9,10,11 \\ 
1H0707-495     &STIS&G140L,G230L & 1122-3155 & 6 \\ 
IRAS13224-3809 &STIS&G140L,G230L & 1117-3148 & 1,6 \\ 
IRAS13349+2438 &FOS &G190H,G270H & 1572-3294 & 4 \\ 
KUG 1031+398   &FOS &G130H,G190H,G270H &1087-3301 & 8,11 \\ 
Mrk110  &STIS&G140M &1194-1250 & 2,11 \\ 
Mrk335  &FOS &G130H,G190H,G270H &1087-3301 & 2,11 \\
        &GHRS&G160M &1221-1257 & \\
Mrk478  &FOS &G130H,G190H,G270H &1087-3277 & 2,5,11 \\
        &STIS&G140M &1194-1300 & \\ 
Mrk486  &FOS &G190H,G270H &1567-3293 & 2,11 \\ 
Mrk493  &FOS &G130H,G190H,G270H,G400H,G570H &1087--6817 & 7,11 \\ 
Mrk766  &STIS&G140L,G230L &1087-3151 & 3,5,7,11 \\ 
NGC4051 &STIS&E140M &1140-1729 & 6,11 \\ 
PG 1211+143 &FOS &G130H,G190H,G270H &1087-3275 & 2,11 \\
            &GHRS&G140L &1190-1477 & \\
            &STIS&G140M &1194-1300 & \\ 
PG 1351+640 &FOS &G130H,G190H,G270H &1087-3301 & 2 \\
            &STIS&G140M &1194-1300 & \\
            &STIS&G230L &1568-3151 & \\ 
PG 1404+226 &FOS &G130H,G190H,G270H,G400H&1087-4780 & 2,11 \\ 
PG 1411+442 &FOS &G130H,G190H,G270H &1087-3276 & 2 \\
            &STIS&G230L &1572-3157 & \\ 
PG 1444+407 &FOS &G130H,G190H &1087-2330 & 3 \\ 
RX J0134-42 &FOS &G130H,G190H,G270H,G400H,G570H&1087--6817 & 5 \\ 
Ton S180 &STIS&G140M &1194-1299 & 5,11,12 \\
         &STIS&G140L,G230L &1120-3160 & \\ 
WPVS007  &FOS &G130H,G190H,G270H,G400H,G570H &1087--6817 & 5,13 \\ 
I Zw1    &FOS &G130H,G190H,G270H &1087-3276 & 2,11 \\ 
         &FOS &G190H,G270H &1568-3295 & \\
         &FOS &G270H &2222-3277 & \\ 
         &GHRS&G160M &1221-1257 & \\ 
II Zw136 &STIS&G140M &1194-1300 & 2 \\ 
         &GHRS&G140L &1153-1739 & \\ 
\enddata
\tablenotetext{a}{ given in \AA} 
\tablenotetext{b}{References to optical emission-line measurements 
that led to their NLS1 classification} 
\tablerefs{ (1) \citet{bol93}; (2) \citet{bor92}; (3) \citet{goo89}; 
(4) \citet{gru98}; (5) \citet{gru99}; (6) \citet{lei99}; (7) \citet{ost85}; 
(8) \citet{puc95}; (9) \citet{sti90}; (10) \citet{gro93}; (11) \citet{ver01}; 
(12) \citet{win92a}; (13) \citet{win92b}}
\end{deluxetable}

\begin{deluxetable}{llccrcr}
\tablewidth{0pt}
\tablecaption{NLS1 Properties \label{tbl-2}}
\tablehead{
\colhead{Object} & \colhead{z\tablenotemark{a}} & 
\colhead{log$L_{\nu}$(1450)\tablenotemark{b}} & 
\colhead{$\alpha$($f_{\nu} \propto \nu^{\alpha}$)} &
\colhead{EW(\ion{C}{4})\tablenotemark{c}} & 
$\Gamma_{\rm ROSAT}$ & Ref\tablenotemark{e}}
\startdata
Ark 564\tablenotemark{d}        & 0.0239 & 28.25 & -1.34$\pm0.01$&   31.6&3.4$\pm$0.1&1\\
1H0707-495\tablenotemark{d}     & 0.0355 & 29.01 & -0.46$\pm0.01$&   20.5&2.3$\pm$0.3&3\\
IRAS13224-3809                  & 0.0580 & 29.03 & -0.44$\pm0.02$&   33.0&4.5$\pm$0.1&1, 3\\
IRAS13349+2438                  & 0.1038 & 28.55 & -3.23$\pm0.06$&   55.3&2.8$\pm$0.1&3\\
KUG 1031+398                    & 0.0432 & 28.10 & -0.99$\pm0.02$&   92.4&4.3$\pm$0.1&1\\
Mrk110                          & 0.0340 &\nodata&\nodata        &\nodata&2.4$\pm$0.1&4\\
Mrk335                          & 0.0260 & 29.23 & -0.64$\pm0.01$&   77.7&2.9$\pm$0.1&3\\
Mrk478                          & 0.0755 & 29.86 & -0.68$\pm0.01$&   39.7&3.1$\pm$0.1&1, 3\\
Mrk486\tablenotemark{d}         & 0.0387 & 27.84 & -2.84$\pm0.09$&   50.0&\nodata&\nodata\\
Mrk493         			& 0.0311 & 28.35 & -0.73$\pm0.01$&  107.5&2.7$\pm$0.2&1\\
Mrk766\tablenotemark{d}         & 0.0127 & 26.13 & -4.41$\pm0.04$&   98.1&2.7$\pm$0.1&1, 3\\
NGC4051\tablenotemark{d}        & 0.00245& 26.38 & -1.86$\pm0.04$&   75.5&2.8$\pm$0.0&3\\
PG 1211+143    			& 0.0804 & 29.75 & -1.14$\pm0.01$&   66.5&3.1$\pm$0.2&3, 4\\
PG 1351+640\tablenotemark{d}    & 0.0875 & 29.65 & -1.35$\pm0.01$&  115.0&2.5$\pm$0.6&4\\
PG 1404+226\tablenotemark{d}    & 0.0935 & 29.22 & -0.91$\pm0.01$&   44.7&4.1$\pm$0.2&3, 4\\
PG 1411+442\tablenotemark{d}    & 0.0875 & 29.68 & -0.98$\pm0.03$&  148.0&3.0$\pm$0.5&4\\
PG 1444+407    			& 0.2640 & 30.55 & -0.81$\pm0.04$&   42.1& \nodata   &\nodata\\
RX J0134-42\tablenotemark{d}    & 0.2360 & 30.31 & -0.12$\pm0.01$&   27.5&7.7$\pm$2.6&2\\
Ton S180       			& 0.0613 & 29.63 & -0.76$\pm0.01$&   33.0&3.0$\pm$0.1&3\\
WPVS007 \tablenotemark{d}       & 0.0280 & 27.78 & -1.41$\pm0.01$&  112.2&9.0$\pm$2.0&2\\
I Zw1          			& 0.0585 & 29.40 & -1.75$\pm0.01$&   50.0&3.1$\pm$0.1&1, 3\\
II Zw136       			& 0.0619 & 29.59 & -0.73$\pm0.04$&   61.2&3.2$\pm$...&4\\
\enddata
\tablenotetext{a}{ calculated from the observed spectra using the 
\ion{C}{4} emission-line; the only exception is Mrk110 for which the 
line is not available, and the redshift is taken from the literature} 
\tablenotetext{b}{to ease comparison with earlier published work, an
$H_{\circ}$ = 50 km s$^{-1}$Mpc$^{-1}$, $q_{\circ} = 0.5$, $\Lambda =
0$ cosmology was adopted.  The values reflect flux measurements which
are corrected for Galactic extinction; typical uncertainties are $\sim$0.05 dex.}
\tablenotetext{c}{rest-frame EWs (in \AA) calculated by fitting a single 
Lorentzian profile to the line} 
\tablenotetext{d}{Emission lines are contaminated by absorption; indicated 
measurements (EWs), obtained from the polynomial interpolated profiles, 
should be regarded as lower limits}
\tablenotetext{e}{References for the $\Gamma_{\rm ROSAT}$ values} 
\tablerefs{ (1) \citet{bol96}; (2) \citet{gru98}; (3) \citet{lei99}; 
(4) \citet{wan96} }
\end{deluxetable}

\begin{deluxetable}{llcccccccc}
\tablewidth{0pt}
\tablecaption{ Emission-line measurements\label{tbl-3}}
\tablehead{
\colhead{} & \colhead{$\lambda_{lab}$\tablenotemark{c}} & 
\colhead{$\lambda_{mean\_rest}$\tablenotemark{d}} & \colhead{$\Delta v$} & 
\colhead{Rel. Flux\tablenotemark{f}} & \colhead{EW\tablenotemark{g}} &  
\colhead{FWHM\tablenotemark{h}} \\
\colhead{Line\tablenotemark{a}} & \colhead{(\AA)} &  \colhead{(\AA)} &
\colhead{(km s$^{-1}$)} & 
\colhead{[100 $\times {\rm F}/{\rm F(Ly}\alpha)$]} & 
\colhead{(\AA)} & \colhead{(km s$^{-1}$)}}
\startdata
\ion{C}{3}$^{\star}$        &1175.5 &1175.5$\pm0.5$&     0$\pm 130$&  0.9$\pm0.2$&  1.1& 1190$\pm$ 690 \\
\ion{Si}{2}                 &1197.4 &1199.3$\pm0.7$&   480$\pm 170$&  0.5$\pm0.2$&  0.6&  790$\pm$ 500 \\
\ion{Si}{3}                 &1206.5 &1206.7$\pm1.9$&    50$\pm 470$&  0.4$\pm0.6$&  0.5&  780$\pm$1350 \\
Ly$\alpha$                  &1215.7 &   \nodata    &   \nodata     &100.0$\pm1.6$&131.8&     \nodata   \\
..narrow                    &1215.7 &1216.3$\pm0.1$&   150$\pm  25$& 69.5$\pm0.9$& 91.6& 1480$\pm$  10 \\
..broad                     &1215.7 &1212.6$\pm1.4$& --760$\pm 340$& 30.6$\pm1.0$& 40.2& 7000$\pm$ 280 \\
\ion{N}{5}                  &1240.8 &1240.0$\pm0.1$& --190$\pm  20$& 13.0$\pm1.5$& 17.6& 2110$\pm$ 130 \\
\ion{Si}{2}$^{\star}$       &1248.4 &1245.0$\pm0.4$& --810$\pm  90$&  3.7$\pm1.0$&  5.0& 2200$\pm$ 400 \\
\ion{Si}{2}                 &1264.8 &1262.9$\pm0.9$& --450$\pm 210$&  3.5$\pm0.4$&  4.8& 3510$\pm$ 350 \\
\ion{O}{1}                  &1303.5 &1305.4$\pm0.3$&   440$\pm  70$&  3.7$\pm0.6$&  5.5& 2060$\pm$ 130 \\
\ion{Si}{2}                 &1309.3 &1309.1$\pm1.5$&  --40$\pm 340$&  1.0$\pm0.7$&  1.5& 2030$\pm$ 280 \\
\ion{C}{2}                  &1335.3 &1336.5$\pm0.3$&   270$\pm  70$&  2.6$\pm0.3$&  4.0& 1930$\pm$ 260 \\
\ion{N}{2}\tablenotemark{b} &1344.6 &1349.0$\pm0.9$&   980$\pm 200$&  1.1$\pm0.3$&  1.7& 2020$\pm$ 430 \\ 
\ion{Si}{4}+\ion{O}{4}]     &1400.0 &   \nodata    &   \nodata     & 10.5$\pm3.2$& 17.2&     \nodata   \\
..\ion{Si}{4}               &1393.7 &1393.7$\pm0.7$&     0$\pm 150$&  3.1$\pm2.2$&  5.1& 1520$\pm$ 600 \\
..\ion{O}{4}]               &1401.4 &1404.6$\pm1.5$&   680$\pm 320$&  5.4$\pm2.3$&  8.9& 2030$\pm$ 440 \\
..\ion{Si}{4}               &1402.7 &1399.9$\pm2.7$& --590$\pm 570$&  2.0$\pm0.7$&  3.3& 1520$\pm$ 690 \\
\ion{N}{4}]                 &1486.5 &1486.8$\pm0.9$&    60$\pm 180$&  0.7$\pm0.3$&  1.2& 2020$\pm$1910 \\
\ion{Si}{2}                 &1530.1 &1536.9$\pm0.2$&  1330$\pm  40$&  4.6$\pm0.8$&  7.9& 2030$\pm$ 280 \\*
\ion{C}{4}                  &1549.5 &   \nodata    &   \nodata     & 44.8$\pm1.7$& 77.1&     \nodata   \\*
..narrow                    &1549.5 &1549.5$\pm0.1$&     0$\pm  20$& 33.3$\pm1.3$& 57.3& 1900$\pm$  60 \\*
..broad                     &1549.5 &1541.6$\pm8.2$&--1520$\pm1580$& 11.5$\pm1.1$& 19.8&12000$\pm$2500 \\
\ion{He}{2}                 &1640.4 &   \nodata    &   \nodata     & 23.3$\pm2.6$& 41.5&     \nodata   \\
..narrow                    &1640.4 &1639.7$\pm0.2$& --120$\pm  40$&  6.4$\pm0.8$& 11.4& 1780$\pm$ 240 \\
..broad                     &1640.4 &1630.4$\pm3.0$&--1820$\pm 540$& 16.9$\pm2.5$& 30.1&11990$\pm$4090 \\
\ion{O}{3}]                 &1663.5 &1665.4$\pm0.2$&   340$\pm  40$&  4.1$\pm1.1$&  6.2& 2500$\pm$ 480 \\
\ion{Al}{2}                 &1670.8 &1680.0$\pm1.5$&  1650$\pm 260$&  1.3$\pm1.0$&  1.9& 2250$\pm$1330 \\
\ion{N}{2}                  &1725.2 &1725.6$\pm0.9$&    70$\pm 150$&  2.2$\pm0.9$&  3.3& 4710$\pm$2100 \\
\ion{N}{3}]                 &1750.5 &1750.5$\pm0.3$&     0$\pm  50$&  2.0$\pm0.2$&  3.0& 1480$\pm$ 170 \\
\ion{Fe}{2} UV191           &1785.4 &1785.3$\pm0.7$&  --20$\pm 110$&  1.2$\pm0.2$&  1.8& 1430$\pm$ 250 \\
\ion{Si}{2}                 &1814.1 &1816.9$\pm0.5$&   460$\pm  80$&  0.6$\pm0.2$&  0.9& 1250$\pm$ 730 \\
\ion{Al}{3}                 &1858.8 &1853.5$\pm1.1$& --850$\pm 170$&  1.3$\pm0.5$&  1.9& 2240$\pm$ 540 \\
\ion{Fe}{3} UV52            &1867.9 &1864.1$\pm0.8$& --610$\pm 120$&  1.7$\pm0.5$&  2.6& 2760$\pm$ 730 \\
\ion{Si}{3}]                &1892.0 &1892.7$\pm0.1$&   110$\pm  20$&  3.6$\pm0.5$&  5.5& 1250$\pm$ 130 \\
\ion{C}{3}                  &1907.9 &   \nodata    &   \nodata     & 21.1$\pm5.7$& 32.2&     \nodata   \\
..narrow                    &1907.9 &1908.2$\pm0.1$&    50$\pm  20$& 10.0$\pm0.6$& 15.3& 1380$\pm$  10 \\
..broad                     &1907.9 &1907.7$\pm4.6$&  --30$\pm 720$& 11.0$\pm5.7$& 16.9&12230$\pm$1110 \\
\ion{N}{2}]                 &2141.4 &2144.0$\pm0.9$&   360$\pm 130$&  0.3$\pm0.2$&   0.4&1040$\pm1140$ \\
\ion{C}{2}]                 &2327.5 &2326.6$\pm0.8$& --120$\pm 100$&  3.7$\pm1.3$&   5.7&2540$\pm 430$ \\
\ion{Fe}{3} UV 47           &2419.3 &2422.8$\pm0.2$&   430$\pm  30$&  3.2$\pm0.5$&   5.1&1560$\pm 200$ \\
\ion{O}{2}                  &2438.8 &2439.3$\pm0.6$&    60$\pm  70$&  0.7$\pm0.2$&   1.2&1060$\pm 250$ \\*
$[$\ion{O}{2}]              &2470.9 &2467.6$\pm0.8$& --400$\pm 100$&  0.9$\pm0.3$&   1.5&1570$\pm 470$ \\*
\ion{C}{1}]\tablenotemark{b}&2478.6 &2484.0$\pm0.1$&   650$\pm  10$&  2.3$\pm0.6$&   3.7&2580$\pm 510$ \\
\ion{Al}{2}]                &2669.9 &2674.7$\pm0.4$&   540$\pm  50$&  0.8$\pm0.6$&   1.3&1050$\pm 430$ \\
\ion{Mg}{2}                 &2797.9 &    \nodata   &   \nodata     & 31.4$\pm6.6$&  58.0&   \nodata    \\
..narrow                    &2797.9 &2801.6$\pm0.3$&   390$\pm  30$& 11.2$\pm2.2$&  20.7&1680$\pm  60$ \\
..broad                     &2797.9 &2798.3$\pm0.4$&    40$\pm  40$& 20.2$\pm6.5$&  37.4&4620$\pm1090$ \\
\ion{O}{3}\tablenotemark{e} &2960.6 &2957.4$\pm1.7$& --320$\pm 170$&  0.3$\pm0.2$&   0.9& 850$\pm 600$ \\
\ion{Fe}{2}\tablenotemark{e}&2964.3 &2966.7$\pm3.6$&   240$\pm 360$&  0.1$\pm0.1$&   0.2& 830$\pm 900$ \\
\ion{O}{4}\tablenotemark{e} &2982.5 &2981.9$\pm2.5$&  --60$\pm 250$&  0.3$\pm0.3$&   0.9& 820$\pm 610$ \\
\ion{Ne}{3}]\tablenotemark{e}&2986.9&2990.0$\pm0.1$&   310$\pm  10$&  0.2$\pm0.1$&   0.5& 830$\pm 680$ \\
\ion{F}{5}\tablenotemark{e} &3109.0 &3109.5$\pm0.4$&    50$\pm  40$&  0.6$\pm0.1$&   2.0& 550$\pm  10$ \\
\ion{O}{3}\tablenotemark{e} &3122.5 &3122.9$\pm1.3$&    40$\pm 120$&  0.8$\pm0.4$&   2.7& 910$\pm 370$ \\
\ion{O}{3}\tablenotemark{e} &3133.7 &3134.2$\pm1.3$&    50$\pm 120$&  1.8$\pm0.4$&   6.0& 940$\pm 210$ \\
\ion{C}{2}\tablenotemark{e} &3166.7 &3165.3$\pm0.6$& --130$\pm  60$&  0.8$\pm0.3$&   2.7& 760$\pm 390$ \\
\ion{He}{1}                 &3188.6 &3190.6$\pm1.7$&   190$\pm 160$&  0.7$\pm0.1$&   1.6& 990$\pm 230$ \\
\ion{O}{4}\tablenotemark{b} &3216.8 &3215.7$\pm0.5$& --100$\pm  50$&  0.5$\pm0.1$&   1.1& 830$\pm 120$ \\
\ion{Ne}{2}\tablenotemark{b}&3230.6 &3230.9$\pm0.8$&    30$\pm  70$&  0.9$\pm0.2$&   2.1& 940$\pm 260$ \\
\ion{Fe}{1} Opt91           &3261.2 &3260.8$\pm0.7$&  --40$\pm  60$&  1.0$\pm0.1$&   2.5& 840$\pm 130$ \\
\ion{Fe}{2} Opt1            &3281.2 &3281.6$\pm0.5$&    40$\pm  40$&  1.2$\pm0.1$&   3.0& 840$\pm  70$ \\
$[$\ion{Fe}{3}]\tablenotemark{b}&3308.5 &3307.8$\pm2.4$&--60$\pm220$& 0.7$\pm0.1$&   1.9& 830$\pm 110$ \\
$[$\ion{Ne}{5}]             &3346.8 &3345.3$\pm0.2$& --130$\pm  20$&  0.5$\pm0.3$&  1.2& 790$\pm 260$ \\*
$[$\ion{Ne}{5}]             &3426.8 &3425.3$\pm0.2$& --130$\pm  20$&  1.2$\pm0.3$&  3.3& 790$\pm 260$ \\*
\ion{Fe}{2}]\tablenotemark{b}&3495.6&3496.2$\pm0.5$&    50$\pm  40$&  1.5$\pm0.1$&  4.4& 880$\pm 100$ \\
$[$\ion{Fe}{7}]             &3587.1 &3584.1$\pm1.0$& --250$\pm  80$&  0.6$\pm0.3$&  1.8&1080$\pm 180$ \\
\ion{O}{2}]\tablenotemark{b}&3621.0 &3620.6$\pm1.0$&  --30$\pm  80$&  0.6$\pm0.1$&  1.8&1280$\pm 310$ \\
\ion{N}{3}]\tablenotemark{b}&3685.9 &3686.3$\pm0.8$&    30$\pm  60$&  0.5$\pm0.3$&  1.5&1170$\pm 110$ \\
$[$\ion{O}{2}]              &3727.1 &3727.7$\pm0.2$&    50$\pm  20$&  1.0$\pm0.1$&  3.3& 610$\pm  70$ \\
$[$\ion{O}{2}]              &3729.8 &3730.3$\pm0.2$&    50$\pm  20$&  0.1$\pm0.1$&  0.2& 610$\pm  70$ \\
$[$\ion{Ne}{2}]             &3745.7 &3749.0$\pm0.1$&   290$\pm  20$&  0.4$\pm0.1$&  1.0& 870$\pm 250$ \\
$[$\ion{Fe}{7}]             &3759.7 &3758.3$\pm0.6$& --110$\pm  60$&  0.5$\pm0.2$&  1.4& 790$\pm 350$ \\
$[$\ion{Fe}{4}]             &3769.7 &3769.4$\pm1.2$&  --30$\pm 100$&  0.3$\pm0.1$&  1.1& 790$\pm 140$ \\
\ion{Fe}{1}                 &3785.3 &3787.2$\pm1.8$&   150$\pm 140$&  1.7$\pm0.1$&  5.6&3620$\pm 190$ \\
$[$\ion{Ne}{3}]             &3870.1 &3868.3$\pm0.3$& --140$\pm  20$&  1.1$\pm0.1$&  4.0& 680$\pm 190$ \\
\ion{He}{1}                 &3889.7 &3890.5$\pm2.5$&    60$\pm 190$&  0.7$\pm0.2$&  2.6& 920$\pm$ 410 \\
\ion{O}{3}]                 &3938.7 &3939.2$\pm1.0$&    40$\pm  70$&  0.4$\pm0.1$&  1.5& 830$\pm 190$ \\
$[$\ion{Ne}{3}]             &3968.9 &3966.9$\pm0.3$& --150$\pm  20$&  0.3$\pm0.1$&  1.2& 660$\pm 190$ \\
\ion{O}{2}]\tablenotemark{b}&3969.6 &3971.3$\pm0.5$&   130$\pm  40$&  1.1$\pm0.1$&  4.0& 830$\pm  70$ \\
$[$\ion{S}{2}]              &4069.7 &4069.5$\pm0.8$&  --20$\pm  60$&  0.2$\pm0.1$&  0.8& 600$\pm 270$ \\
$[$\ion{S}{2}]              &4077.5 &4077.3$\pm0.7$&  --20$\pm  50$&  0.1$\pm0.1$&  0.3& 600$\pm 260$ \\
H$\delta$                   &4102.9 &4105.6$\pm0.7$&   200$\pm  50$&  3.7$\pm0.1$& 14.9&2240$\pm 160$ \\
\ion{Fe}{3}]\tablenotemark{b}&4180.0&4181.7$\pm0.8$&   120$\pm  60$&  1.4$\pm0.1$&  5.8&1220$\pm 220$ \\
\ion{Fe}{2}                 &4234.3 &4239.2$\pm1.0$&   350$\pm  70$&  1.3$\pm0.1$&  5.7&1010$\pm 150$ \\
H$\gamma$                   &4341.6 &4344.7$\pm0.2$&   210$\pm  10$&  7.3$\pm0.2$& 26.3&1070$\pm  50$ \\
$[$\ion{O}{3}]              &4364.4 &4367.4$\pm0.1$&   210$\pm  10$&  0.7$\pm0.1$&  2.4& 270$\pm  60$ \\*
\ion{He}{1}\tablenotemark{b}&4472.7 &4477.2$\pm3.1$&   300$\pm 200$&  0.6$\pm0.2$&  2.2&1270$\pm 280$ \\*
\ion{He}{2}                 &4687.0 &4688.6$\pm0.9$&   100$\pm  50$&  4.9$\pm0.1$& 19.0&2890$\pm  80$ \\
H$\beta$                    &4862.6 &   \nodata    &     \nodata   & 20.5$\pm1.1$& 81.5& \nodata      \\
..narrow                    &4862.6 &4864.6$\pm0.1$&   120$\pm  10$& 17.5$\pm0.3$& 69.6&1020$\pm  10$ \\
..broad                     &4862.6 &4864.5$\pm9.9$&   120$\pm 610$&  3.0$\pm1.1$& 11.9&6890$\pm 200$ \\
$[$\ion{O}{3}]              &4960.2 &4963.4$\pm0.1$&    40$\pm  10$& 2.3$\pm0.1$&   9.0& 260$\pm  10$ \\
$[$\ion{O}{3}]              &5008.2 &5008.5$\pm0.1$&    40$\pm  10$& 8.0$\pm0.2$&  31.3& 260$\pm  10$ \\
$[$\ion{Fe}{7}]             &5159.8 &5162.1$\pm1.4$& --130$\pm 130$& 0.9$\pm0.3$&   3.8&1080$\pm 150$ \\
$[$\ion{Fe}{6}]             &5177.5 &5180.2$\pm1.9$& --240$\pm 190$& 1.2$\pm0.3$&   4.8& 860$\pm 250$ \\
$[$\ion{N}{1}]              &5200.5 &5203.5$\pm2.9$&   130$\pm 280$& 0.8$\pm0.8$&   3.3& 870$\pm 820$ \\
$[$\ion{Fe}{7}]             &5277.3 &5278.3$\pm1.4$& --130$\pm 130$& 1.8$\pm0.3$&   7.4&1080$\pm 150$ \\
\ion{O}{4}\tablenotemark{b} &5318.7 &5318.2$\pm0.7$&  --30$\pm  40$& 1.3$\pm0.2$&   5.3& 750$\pm 120$ \\
\ion{He}{1}                 &5877.3 &   \nodata    &   \nodata     &  2.3$\pm0.1$& 11.1&   \nodata   \\
..narrow                    &5877.3 &5878.4$\pm1.1$&    60$\pm  50$&  1.5$\pm0.1$&  7.5& 970$\pm  70$ \\
..broad                     &5877.3 &5881.4$\pm9.1$&   210$\pm 460$&  0.7$\pm0.1$&  3.6&5010$\pm 550$ \\
$[$\ion{Fe}{7}]             &6087.9 &6088.5$\pm1.3$&    30$\pm  60$&  0.3$\pm0.5$&  1.4& 870$\pm 340$ \\
$[$\ion{Fe}{5}]             &6088.5 &6088.8$\pm1.5$&    20$\pm  70$&  0.4$\pm0.6$&  2.0& 870$\pm 290$ \\
$[$\ion{O}{1}]              &6302.1 &6304.5$\pm0.8$&   110$\pm  40$&  1.2$\pm0.1$&  6.4&1150$\pm 150$ \\
$[$\ion{O}{1}]              &6365.5 &6367.5$\pm0.5$&    90$\pm  20$&  0.4$\pm0.1$&  2.2&1140$\pm 150$ \\
$[$\ion{Fe}{10}]            &6376.3 &6373.5$\pm0.6$& --130$\pm  30$&  1.2$\pm0.1$&  6.5& 810$\pm  90$ \\
$[$\ion{N}{2}]              &6548.8 &6549.5$\pm0.9$&    30$\pm  40$&  1.8$\pm0.6$& 10.1& 760$\pm  70$ \\
H$\alpha$                   &6564.6 &   \nodata    &   \nodata     & 61.0$\pm1.4$&344.2& \nodata      \\*
..narrow                    &6564.6 &6565.9$\pm0.1$&    60$\pm   5$& 57.8$\pm1.0$&325.8& 760$\pm  20$ \\*
..broad                     &6564.6 &6566.7$\pm0.4$&    90$\pm  20$&  3.3$\pm1.2$& 18.4&5500$\pm1140$ \\
$[$\ion{N}{2}]              &6585.3 &6585.9$\pm0.9$&    30$\pm  40$&  6.1$\pm0.6$& 34.3& 760$\pm  70$ \\
\enddata       								       
\tablenotetext{a}{Emission lines are measured from the average 
composite spectrum unless otherwise noted}
\tablenotetext{b}{Uncertain identifications}
\tablenotetext{c}{vacuum wavelengths, both below and above 2200\AA}
\tablenotetext{d}{given by the centroid of the Lorentzian fit}
\tablenotetext{f}{The line fluxes are normalized to the Ly$\alpha$ flux to ease
comparison with published results for other AGN/quasar samples}
\tablenotetext{e}{Lines measured from the median composite}
\tablenotetext{g}{The errors in EWs are not quoted as the relative errors 
are the same as for the fluxes}
\tablenotetext{h}{Intrinsic width of the emission line; an instrumental 
broadening of 230 km s$^{-1}$, the approximate FOS spectral resolution, 
high-resolution gratings, was assumed}
\end{deluxetable}

\begin{deluxetable}{lccc}
\tablewidth{0pt}
\tablecaption{ Proportions of PCs in the \ion{C}{4} emission region 
\label{tbl-4}}
\tablehead{
\colhead{Component} & \colhead{NLS1} & \colhead{$z \ga 4$ QSO} & 
\colhead{Combined sample} \\
\colhead {}  & \colhead{(21 obj.)} & \colhead{(44 obj.)} & \colhead{(65 obj.)} }
\startdata
PC1      & 0.358 & 0.272 & 0.296\\ 
PC2      & 0.139 & 0.077 & 0.091\\ 
PC3      & 0.097 & 0.071 & 0.061\\
PC4      & 0.064 & 0.053 & 0.051\\
PC5      & 0.046 & 0.044 & 0.041\\
\enddata   
\end{deluxetable}

\end{document}